\documentclass[fleqn,usenatbib]{mnras}

\usepackage{newtxtext,newtxmath}
\usepackage[T1]{fontenc}
\usepackage{graphicx}	
\usepackage{amsmath}	

\DeclareRobustCommand{\VAN}[3]{#2}
\let\VANthebibliography\thebibliography
\def\thebibliography{\DeclareRobustCommand{\VAN}[3]{##3}\VANthebibliography}

\title[D$_2$CO in the IRAS\,4A outflow cavities]{FAUST XIX. D$_2$CO in the outflow cavities of NGC\,1333 IRAS\,4A: recovering the physical structure of its original prestellar core}

\author[Chahine et al.]{
    Layal Chahine,$^{1}$\thanks{E-mail: layal.chahine@univ-grenoble-alpes.fr}
    Cecilia Ceccarelli,$^{1}$\thanks{E-mail: cecilia.ceccarelli@univ-grenoble-alpes.fr}
    Marta De Simone,$^{2}$
    Claire J. Chandler,$^3$
    Claudio Codella,$^{4}$
    \and
    Linda Podio,$^{4}$
    Ana L\'opez-Sepulcre,$^{1,5}$
    Brian Svoboda,$^{3}$
    Giovanni Sabatini,$^{4}$
    Nami Sakai,$^6$
    Laurent Loinard,$^{7}$
    \and
    Charlotte Vastel,$^8$
    Nadia Balucani,$^{9}$
    Albert Rimola$^{10}$
    Piero Ugliengo,$^{11}$
    Yuri Aikawa,$^{12}$ 
    Eleonora Bianchi,$^{4,13}$
    \and
    Mathilde Bouvier,$^{14}$
    Paola Caselli,$^{15}$
    Steven Charnley,$^{16}$
    Nicol\'as Cuello,$^{1}$
    Tomoyuki Hanawa,$^{17}$
    \and
    Doug Johnstone,$^{18,19}$
    Maria Jos\'e Maureira,$^{15}$
    Francois M\'enard,$^{1}$
    Yancy Shirley,$^{20}$
    Leonardo Testi,$^{21,4}$
    \and
    Satoshi Yamamoto$^{22}$
\\\\
$^1$Université Grenoble Alpes, CNRS, IPAG, 38000 Grenoble, France\\
$^{2}$ESO, Karl Schwarzchild Srt. 2, 85478 Garching bei München, Germany\\
$^3$National Radio Astronomy Observatory, PO Box O, Socorro, NM 87801, USA\\
$^4$INAF, Osservatorio Astrofisico di Arcetri, Largo E. Fermi 5, 50125 Firenze, Italy\\
$^5$Institut de Radioastronomie Millim\'etrique (IRAM), 300 rue de la Piscine, 38406 Saint-Martin-d’Hères, France\\
$^6$RIKEN Cluster for Pioneering Research, 2-1, Hirosawa, Wako-shi, Saitama 351-0198, Japan\\
$^7$Instituto de Radioastronom\'ia y Astrof\'isica, Universidad Nacional Aut\'onoma de M\'exico, A.P. 3-72 (Xangari), 8701, Morelia, M\'exico\\
$^8$IRAP, Université de Toulouse, CNRS, UPS, CNES, 31400 Toulouse, France\\
$^{9}$Dipartimento di Chimica, Biologia e Biotecnologie, Università di Perugia, 06123 Perugia, Italy\\
$^{10}$Departament de Química, Universitat Autonoma de Barcelona, 08193 Bellaterra, Spain\\
$^{11}$Dipartimento di Chimica and Nanostructured Interfaces and Surfaces (NIS) Centre, Universit\`{a} degli Studi di Torino, via P. Giuria 7, 10125, Torino, Italy\\
$^{12}$Department of Astronomy, The University of Tokyo, 7-3-1 Hongo, Bunkyo-ku, Tokyo 113-0033, Japan\\
$^{13}$Excellence Cluster ORIGINS, Boltzmannstrasse 2, D-85748 Garching bei München, Germany\\
$^{14}$Leiden Observatory, Leiden University, P.O. Box 9513, 2300 RA Leiden, The Netherlands\\
$^{15}$Center for Astrochemical Studies, Max-Planck-Institut für extraterrestrische Physik (MPE), Gießenbachstr. 1, D-85741 Garching, Germany\\
$^{16}$Astrochemistry Laboratory, Code 691, NASA Goddard Space Flight Center, 8800 Greenbelt Road, Greenbelt, MD 20771, USA\\
$^{17}$Center for Frontier Science, Chiba University, 1-33 Yayoi-cho, Inage-ku, Chiba 263-8522, Japan\\
$^{18}$NRC Herzberg Astronomy and Astrophysics, 5071 West Saanich
Road, Victoria, BC, V9E 2E7, Canada\\
$^{19}$Department of Physics and Astronomy, University of Victoria, Vic-
toria, BC, V8P 5C2, Canada\\
$^{20}$Steward Observatory, 933 N Cherry Ave., Tucson, AZ 85721 USA\\
$^{21}$Dipartimento di Fisica e Astronomia “Augusto Righi” Viale Berti Pichat 6/2, Bologna, Italy\\
$^{22}$SOKENDAI, Shonan Village, Hayama, Kanagawa 240-0193, Japan
}

\date{Accepted 2024 August 27. Received 2024 August 17; in original form 2024 July 16}

\pubyear{2024}

\begin{document}
\label{firstpage}
\pagerange{\pageref{firstpage}--\pageref{lastpage}}
\maketitle

\begin{abstract}
%
%
Molecular deuteration is a powerful diagnostic tool for probing the physical conditions and chemical processes in astrophysical environments. 
In this work, we focus on formaldehyde deuteration in the protobinary system NGC\,1333 IRAS\,4A, located in the Perseus molecular cloud. 
Using high-resolution ($\sim$\,100\,au) ALMA observations, we investigate the [D$_2$CO]/[HDCO] ratio along the cavity walls of the outflows emanating from IRAS\,4A1. 
Our analysis reveals a consistent decrease in the deuteration ratio (from $\sim$\,60-20\% to $\sim$\,10\%) with increasing distance from the protostar (from $\sim$\,2000\,au to $\sim$\,4000\,au). 
Given the large measured [D$_2$CO]/[HDCO], both HDCO and D$_2$CO are likely injected by the shocks along the cavity walls into the gas-phase from the dust mantles, formed in the previous prestellar phase.
We propose that the observed [D$_2$CO]/[HDCO] decrease is due to the density profile of the prestellar core from which NGC\,1333 IRAS\,4A was born. 
When considering the chemical processes at the base of formaldehyde deuteration, the  IRAS\,4A's prestellar precursor had a predominantly flat density profile within 3000\,au and a decrease of density beyond this radius. 
\end{abstract}

\begin{keywords}
ISM: abundances -- ISM: jets and outflows -- stars: formation -- stars: low-mass -- ISM: individual object: NGC\,1333 IRAS\,4A
\end{keywords}


\section{Introduction} \label{sec:intro}

Since the discovery of the doubly deuterated isotopologue of formaldehyde, D$_2$CO, towards the Orion hot core \citep{Turner1990-D2CO} and the solar-like Hot Corino of IRAS16293-2422 \citep{Ceccarelli1998-D2CO} the [D$_2$CO]/[H$_2$CO] abundance ratio has been used to assess the physical conditions at the time of their formation.
In Orion hot core, [D$_2$CO]/[H$_2$CO] is about 0.3\% while in IRAS16293-2422 it is $\sim$10\%.
Since the cosmic D/H elemental abundance is (2.53$\pm$0.03)$\times 10^{-5}$ \citep{Cooke2018-cosmicD}, even assuming that some deuterium has been used in Milky Way stellar nucleosynthesis, [D$_2$CO]/[H$_2$CO] should be of the order of $10^{-10}$ if the distribution of H and D atoms in formaldehyde were statistical, many orders of magnitude lower than that observed.

The (only known) interstellar process that can enhance this ratio this much is so-called molecular deuteration, which is basically due to the low temperature of the environment where D$_2$CO is formed \citep[e.g.][]{Turner1990-D2CO, Ceccarelli1998-D2CO, Roberts2003-multiplyD, Ceccarelli2014-PP6, Nomura2023-PP7}.
Thus, the very detection of D$_2$CO is already a hallmark that the species is formed in cold gas, despite its being observed in a hot core and hot corino.
The only way to reconcile this apparent contradiction is that D$_2$CO was formed in the previous prestellar cold phase and remained iced on the grain mantles until they sublimated in the hot core/corinos, injecting D$_2$CO (and H$_2$CO) into the gas, where it is detected.
The factor of 30 difference in the [D$_2$CO]/[H$_2$CO] abundance ratio of the Orion hot core and IRAS16293-2422 hot corino points to a difference of the physical conditions in the prestellar precursor from which the two protostars were born, with the latter being colder than the former \citep[see, e.g. the discussion in][]{Ceccarelli1998-D2CO}.

\begin{table*}
    \centering
    \caption{Spectral parameters of the HDCO and D$_{2}$CO lines observed towards IRAS\,4A.}
    \begin{tabular}{cccccccccc}
    
         \hline
       $\mathrm{Transition}$  & $\mathrm{\nu \,} ^{(a)} $ & $\mathrm{E_{up} \,} ^{(a)}$ & $\mathrm{A_{ij} \,} ^{(a)}$ &  $\mathrm{g_{up}\,} ^{(a)}$ & Bandwidth & $\mathrm{dV} $ & FWHM FoV & $\mathrm{Beam \, (PA)}$ & $\mathrm{Chan. \, rms} $  \\
         &  (MHz) & (K) & (10$^{-4}$\,s$^{-1}$) & & (MHz) & (km s$\rm ^{-1}$) & $\arcsec$& $\arcsec$ $\times$ $\arcsec$ ($\rm ^{\circ}$) &
         (mJy beam$\rm ^{-1}$) \\ \hline
          
         HDCO(4$_{1,4}$--3$_{1,3}$) & 246924.60 & 37.6 & 3.98 & 9 & 1875 & 1.19 & 25.5 & 0.37 $\times$ 0.27 (-17.4) & 0.40 \\
         HDCO(4$_{2,2}$--3$_{2,1}$) & 259034.91 & 62.9 & 3.66 & 9 & 58.6 & 0.14 & 24.3 & 0.36 $\times$ 0.25 (-19.8) & 1.13 \\
         D$_{2}$CO(4$_{0,4}$--3$_{0,3}$) & 231410.23 & 27.9 & 3.47 & 18 & 58.6 & 0.17 & 27.2 & 0.45 $\times$ 0.37 (-5.8) & 0.81  \\
         \hline 
    \end{tabular}
    
    \label{tab:line-params}
     \noindent
$^{(a)}$ Frequencies and spectroscopic parameters have been extracted from \cite{Bocquet1999} for both HDCO and D$_2$CO.
\end{table*}

After the discovery of abundant D$_2$CO in the IRAS16293-2422 hot corino, several studies of molecular deuteration focused on the regions where low-mass stars are forming \cite[e.g.,][]{Roueff2000, Loinard2001, Ceccarelli2002-D2COinL1689N, vanderTak2002-ND3, Hatchell2003, Liu2011, Bergman2011-d2co, Taquet2015, Jorgensen2016, Bianchi2017-SVS13deuteration, Jorgensen2018, Manigand2020-i16293a, vanGelder2020, Redaelli2019, Mercimek2022} and where the molecular deuteration is much higher with respect to high-mass star-forming regions \citep{Loinard2002,Fontani2011-highmassDeut,Rivilla2020, Sabatini2020-H2D+, Zahorecz2021-d2coHighMass, Sabatini2023-CR, Pazukhin2023-Dhighhass}.
After an initial surprise for the large measured abundance of these multi-deuterated molecules, it soon became clear that the depletion of CO is another major actor in the molecular deuteration enhancement, as shown by the observed correlation between [D$_2$CO]/[H$_2$CO] and the CO depletion in a sample of prestellar cores \citep{Bacmann2003-COdepD2CO} (the reason for that will be discussed in Sec. \ref{sec:discussion}).
To be noticed, CO depletion is itself enhanced in low temperature and high density environments \citep[e.g.,][]{Caselli1999-COdepl, Bacmann2002-COdepletion, Crapsi2005}.

Given its high diagnostic power, the [D$_2$CO]/[H$_2$CO] abundance ratio has been studied in a variety of objects, from prestellar cores to high-mass hot cores, and solar-like and brown dwarf hot corinos \citep[e.g.][]{Ceccarelli2001-extendedD2CO, Loinard2002, Bacmann2003-COdepD2CO, Parise2006, Roberts2007-D2COobs, Bergman2011-d2co, Zahorecz2021-d2coHighMass, riaz2022-D2COinBDs}.
In cold prestellar cores or around the extended cold envelopes of Class 0 protostars the enhanced [D$_2$CO]/[H$_2$CO] abundance ratio could be a present-day chemical product (i.e. both species are formed in the gas-phase) \citep[e.g.][]{Roueff2007-warmDeuteration, Bergman2011-d2co}.
However, in hot cores and corinos, [D$_2$CO]/[H$_2$CO] is predominantly set by the chemistry that occurred in their previous prestellar phase.
Thus, [D$_2$CO]/[H$_2$CO] allows us to recover the physical structure of their precursors.

In addition, it is essential to consider the [D$_2$CO]/[HDCO] ratio when studying formaldehyde deuteration. The relationship between the abundances of HDCO and D$_2$CO can indicate whether these molecules are primarily formed on grain surfaces or through gas-phase reactions \citep[e.g. ][]{Turner1990-D2CO, Charnley1997-Dmethanol, Ceccarelli1998-D2CO}. This ratio has been utilised in various studies to determine the formation mechanism \citep[e.g.][]{Zahorecz2017, Zahorecz2021-d2coHighMass, riaz2022-D2COinBDs}, including studies on hot corinos, discs and extended structures around low-mass protostellar sources \citep[e.g.][]{Parise2006, Persson2018, Evans2023, Podio2024}, which suggest that these species are products of grain surface chemistry. Moreover, if the [D$_2$CO]/[HDCO] ratio is not altered by gas-phase reactions, it reflects the conditions from the prestellar core phase, thus preserving information about the physical structure of the original molecular cloud.

So far, the only studies of molecular deuteration as a function of the distance from the centre of the core have been obtained in very few prestellar objects, with L1544 representing the best studied of them \cite[e.g.][]{Caselli2003-h2dp, Chacon2019-Dmethanol}.
While these studies help us to understand the typical prestellar core physical and chemical structure, by definition, they cannot provide us with a direct link between them and the future protostar.
In this Letter, we present a new method to reach this goal.
We measured [D$_2$CO]/[HDCO] along the cavities of the outflows emanating from a young protostar, NGC\,1333 IRAS\,4A to reconstruct the physical conditions of its prestellar core in the inner $\sim$4000 au (in radius).
To this end, we exploited observations of HDCO and D$_2$CO line emission obtained within the ALMA Large Program \textit{Fifty AU STudy of the chemistry in the disc/envelope system of solar-like protostars} (FAUST; 2018.1.01205.L, PI: S. Yamamoto; \citealt{Codella2021}).

\section{Source background} \label{sec:source}

The protobinary system NGC\,1333 IRAS\,4A, situated in the Perseus molecular cloud approximately 300 pc away \citep{Zucker2018}, has captivated researchers due to its unique characteristics and intricate chemistry and dynamics. 
It harbours two prominent protostars, namely IRAS\,4A1 (referred to as 4A1) and IRAS\,4A2 (referred to as 4A2). 
These protostars are encased within a shared envelope with a mass of roughly 8 M$_{\odot}$ \citep{Maury2019} and have a combined bolometric luminosity of approximately 9 L$_{\odot}$ \citep{Kristensen2012}. 
One distinguishing feature of this system is its intricate outflow phenomena, extensively investigated using a variety of molecular tracers, including interstellar complex organic molecules \cite[e.g.][]{Blake1995, Lefloch1998a, Girart1999, DiFrancesco2001, Choi2001, Choi2005, Santangelo2015, Ching2016, Su2019, Taquet2020, Desimone2020-I4outflow, Podio2021, Chuang2021}. 
The outflow emission within the system is notably complex, with overlaps observed particularly at the northern region of the protostars, posing challenges for comprehensive characterisation. 
A recent study of the large scale emission ($\sim$\,7000\,au) proposed the presence of two outflow systems associated with each protostar, with 4A1 exhibiting three outflow cavities and 4A2 displaying four \citep{Chahine2024}.

Despite the extensive research on NGC\,1333 IRAS\,4A, studies on its deuterium fractionation remain limited. 
Previous investigations focused on the envelope, examining formaldehyde, ammonia, methanol, and water through single-dish observations \citep{Loinard2002, vanderTak2002-ND3, Parise2006, Coutens2013, Koumpia2017}. 
Notably, the deuteration of formaldehyde, the focus of this paper, was estimated in different works. 
\cite{Loinard2002} reported a [D$_2$CO]/[H$_2$CO] ratio of $\sim$\,7.3\%. Based on several transitions, \cite{Parise2006} reported a [D$_2$CO]/[HDCO] ratio of $62.0^{+33.0}_{-26.0}$ \% within the inner 3000\,au, while \cite{Koumpia2017} reported a ratio of $\sim$\,100\% based on a single D$_2$CO line.
Similarly, studies of the hot corino component explored the deuteration of water, ammonia, methanol and methyl cyanide \citep{Coutens2013, Yamato2022-DammoniaIRAS4A, Taquet2013-obs, Persson2014, Taquet2019, Yang2021-peaches}. 
However, investigations specifically targeting deuterium fractionation within the outflow component are lacking. 
Furthermore, the intricate nature of the system, with its multiple outflows, underscores the importance of employing high-resolution interferometric observations to fully disentangle the various components.


\begin{figure*}
    \centering
    \includegraphics[width=0.98\textwidth]{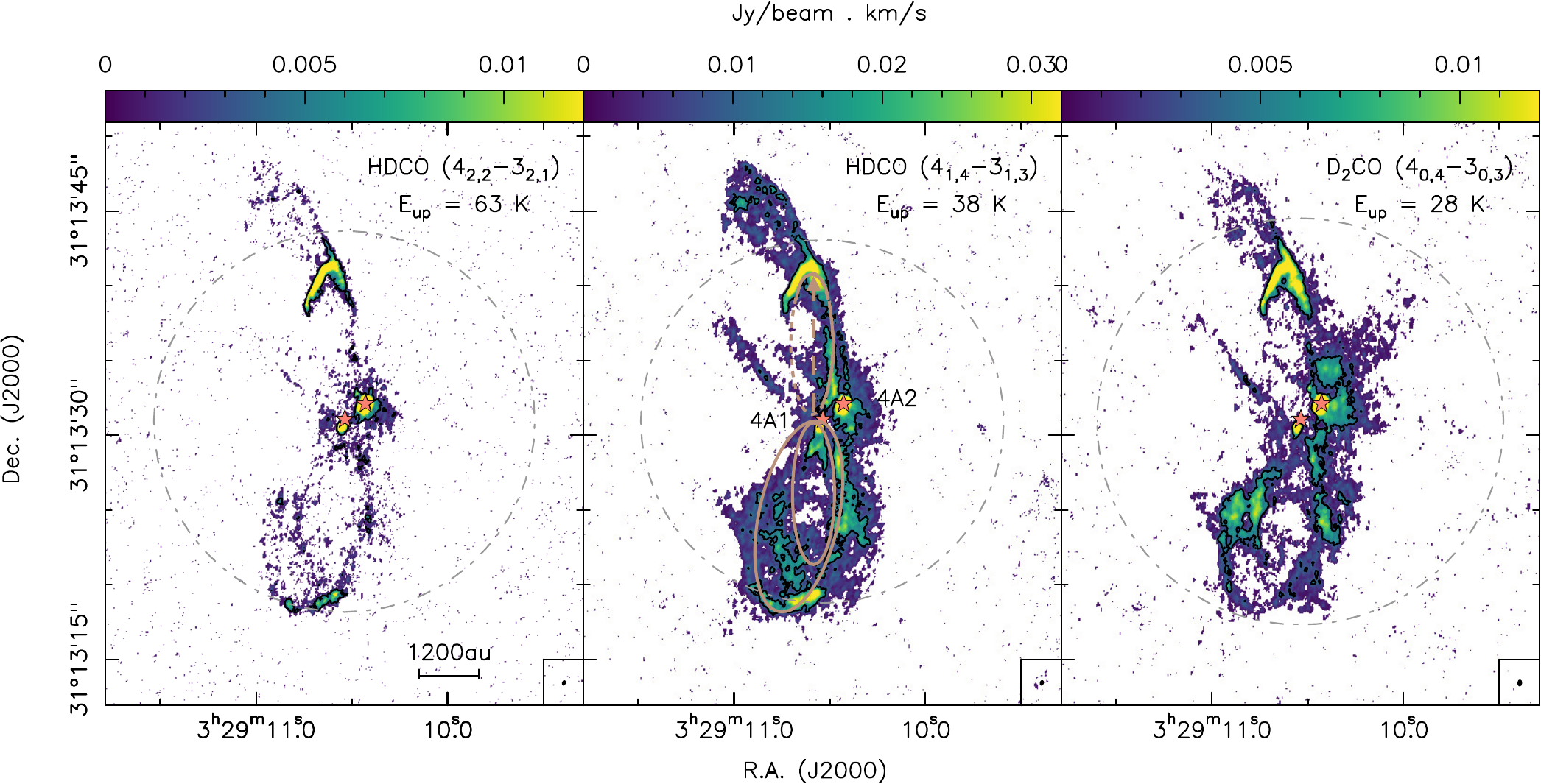}
    \caption{Integrated intensity (moment\,0) maps of HDCO(4$_{2,2}$--3$_{2,1}$), HDCO(4$_{1,4}$--3$_{1,3}$), and D$_{2}$CO(4$_{0,4}$--3$_{0,3}$) towards the NGC\,1333 IRAS\,4A system. A threshold of 3$\sigma$ was applied for all the maps. The HDCO(4$_{2,2}$--3$_{2,1}$) emission is integrated between 2.0 and 11.5\,km\,s$^{-1}$ and the contours are at 5$\sigma$ with $\sigma$\,=\,1.13\,mJy\,beam$^{-1}$\,km\,s$^{-1}$. The HDCO(4$_{1,4}$--3$_{1,3}$) emission is integrated between -3.6 and 20.0\,km\,s$^{-1}$ and the contours are at 30$\sigma$ with $\sigma$\,=\,0.40\,mJy\,beam$^{-1}$\,km\,s$^{-1}$. The D$_{2}$CO emission is integrated between 3.4 and 10.6\,km\,s$^{-1}$ and the contours are at 5$\sigma$ with $\sigma$\,=\,0.81\,mJy\,beam$^{-1}$\,km\,s$^{-1}$. The colour scales are shown at the top of each panel.
    The protostars are depicted with salmon stars. The solid ellipses in the middle panel represent the IRAS\,4A1 outflow cavities. The synthesised beams of the molecules are depicted in the lower right corner.}
    \label{fig:hdco-d2co-mom}
\end{figure*}

\section{Observations \& Results} \label{sec:obs-results}

\subsection{Observations} \label{subsec:observations}

The NGC\,1333 IRAS\,4A system was observed at 1.2\,mm with ALMA during its Cycle 6 operations, between October 2018 and September 2019, as part of the Large Program FAUST. 
The observations were performed in Band 6 using the 12-m array. The baselines for the 12-m array were between 15.1\,m and 3.6\,km, probing angular scales from 0$\farcs$03 ($\sim$9\,au) to 8$\farcs$2 ($\sim$2400\,au).
The observations were centred at R.A. (J2000)\,=\, 03$^{\rm{h}}29^{\rm{m}}10^{\rm{s}}$.539, Dec. (J2000)$ \,= 31^{\circ}13'30''.92$ and the systemic velocity was set to V$_{\rm{lsr}}$\,=\,7\,km\,s$^{-1}$. 
Several spectral windows (spw) were placed within the spectral range 216–234 GHz and 243–262 GHz. In this work, we focus on two narrow spws covering the following lines: D$_{2}$CO(4$_{0,4}$--3$_{0,3}$) and HDCO(4$_{2,2}$--3$_{2,1}$), and one large spw covering HDCO(4$_{1,4}$--3$_{1,3}$). The spectroscopic parameters of each line are reported in Table \ref{tab:line-params}, together with the spw bandwidth and channel width.
The quasar J0237+2848 was used for bandpass and flux calibration, while J0336+3218 and J0328+3139 were used for phase calibration. The absolute flux calibration uncertainty is estimated to be <15\%. The data calibration was performed using the standard ALMA calibration pipeline with the Common Astronomy Software Applications package 5.6.1-8 (CASA\footnote{\url{https://casa.nrao.edu/}}, \citealp{CasaTeam2022}). An additional calibration routine\footnote{\url{https://help.almascience.org/kb/articles/what-are-the-amplitude-calibration-issues-caused-by-alma-s} \\ \url{-normalization-strategy}} has been used to correct for the $T_{sys}$ and for spectral data normalisation. The continuum was generated using careful identification of line-free channels and self-calibration was performed. The continuum model was then subtracted from the line channels to produce continuum-subtracted visibilities (See \citep{Chahine2024} for details). The visibility data from the 12-m arrays were then combined to produce the continuum-subtracted line cubes. The resulting continuum-subtracted cubes were cleaned using H\"{o}gbom's algorithm \citep{Hogbom1974} and normal weighting in the IRAM-GILDAS software package\footnote{http://www.iram.fr/IRAMFR/GILDAS/}, and were corrected for the primary beam attenuation. The data analysis was performed using the same package. The resulting beam size and rms are summarised in Table \ref{tab:line-params}.  In addition, to ensure the accuracy of flux measurements near the edge of the field of view in the original cubes, each cube was also imaged using natural weighting and a 200 k$\lambda$ taper using the task  {\tt tclean} CASA.

\renewcommand{\arraystretch}{1.8}
\begin{table*}
    \centering
    \caption{D$_2$/D ratios obtained at various positions along the southern cavity of IRAS 4A1.}
    \begin{tabular}{ccccccc}
    \hline
        \hline
        Position Name & Position Offset from 4A1  & Projected Distance from 4A1 & $\mathrm{T}_{\mathrm{ex}}\,^{\rm{(a)}}$ & $\mathrm{I}_{\mathrm{HDCO}}$ & $\mathrm{I}_{\mathrm{D_2CO}}$ & D$_{2}$/D\,$^{\rm{(b)}}$ \\
         & ($''$, $''$)& (au) &(K) & (K km s$^{-1}$) & (K km s$^{-1}$) & (\%) \\
         \hline  

        N & (2$\farcs$4, 7$\farcs$1) & 2240 & $36.3 \pm 12.6$ & $3.4 \pm 0.3 $ & $1.5 \pm 0.2$ & $30.3^{+8.5}_{-6.5}$ \\
        A & (4$\farcs$2, -6$\farcs$0) & 2190 & $25.2 \pm 6.2$ & $2.6 \pm 0.2$ & $1.7 \pm 0.1$ & $52.8^{+10.6}_{-9.7}$ \\
        B & (2$\farcs$8, -5$\farcs$8) & 1930 & $25.2 \pm 6.2$ & $4.1 \pm 0.4$ & $2.1 \pm 0.2$ & $41.5^{+8.9}_{-7.9}$ \\
        C & (-1$\farcs$6, -7$\farcs$3) & 2230 & $25.1 \pm 5.8$ & $5.3 \pm 0.3$ & $1.6 \pm 0.1$ & $24.4^{+4.7}_{-4.4}$ \\
        D & (4$\farcs$5, -11$\farcs$7) & 3750 & $23.4 \pm 5.8$ & $4.8 \pm 0.4$ & $0.7 \pm 0.1$ & $11.9^{+5.0}_{-3.4}$ \\
        E & (3$\farcs$5, -12$\farcs$3) & 3820 & $26.7 \pm 6.0$ & $8.6 \pm 0.6$ & $1.0 \pm 0.2$ & $9.6^{+2.5}_{-2.0}$ \\
        
         \hline  
    \end{tabular}
    \label{tab:deuteration-results}
    \footnotesize
    \newline
    \textbf{Notes:} $^{\rm{(a)}}$ T$_{\mathrm{ex}}$ values are determined through RD analysis of HDCO. \\
       $^{\rm{(b)}}$ For the calculation of the D$_{2}$/D ratio, T$_{\mathrm{ex}}$ is assumed to be $25.0 \pm 6.0$ K, derived from the averaged values obtained at various southern points. The error calculation incorporates both intensity and T$_{\mathrm{ex}}$ errors.
\end{table*}

\subsection{HDCO and D$_{2}$CO maps} \label{subsec:maps}

We imaged the emission lines of HDCO(4$_{1,4}$--3$_{1,3}$), HDCO(4$_{2,2}$--3$_{2,1}$), and D$_2$CO(4$_{0,4}$--3$_{0,3}$) towards IRAS\,4A, with the resulting moment\,0 maps depicted in Fig. \ref{fig:hdco-d2co-mom}. The emission from HDCO(4$_{1,4}$--3$_{1,3}$) has been previously discussed in a prior publication \citep{Chahine2024}, where it was observed to trace the outflow cavity walls in the vicinity of the IRAS\,4A system.

Towards the south of the protostars, we discern two cavities driven by IRAS\,4A1: one extending with a position angle (PA) of approximately -12$^{\circ}$ to the east, and another nested within the former and oriented in a north-south direction. Additionally, part of the east cavity wall of IRAS\,4A2 is also evident. To the north, there is an overlapping of emission from outflows driven by 4A1 and 4A2. However, our previous work suggests that HDCO traces the northern counterpart of the north-south cavity driven by 4A1, as well as another cavity associated with IRAS\,4A2. Notably, the emission from HDCO(4$_{2,2}$--3$_{2,1}$) follows a similar pattern, albeit weaker, owing to its higher excitation energy.

D$_2$CO also delineates the cavity walls, marking the first observation of such features using this molecule. The emission pattern of D$_2$CO closely mirrors that of HDCO(4$_{1,4}$--3$_{1,3}$), tracing the outflows driven by 4A1 towards the south and both the cavity walls of 4A1 and 4A2 towards the north.

In the analysis here, we focus on deriving the deuteration ratio using HDCO and D$_2$CO. Even though H$_2$CO is available in our dataset, we limit the analysis to HDCO and D$_2$CO due to the complex H$_2$CO emission profile and its tracing of multiple components, as determined in our previous work.


\subsection{Intensity ratio and deuterium fractionation} \label{subsec:Deuteration}

\begin{figure}
    \centering
    \includegraphics[width=0.46\textwidth]{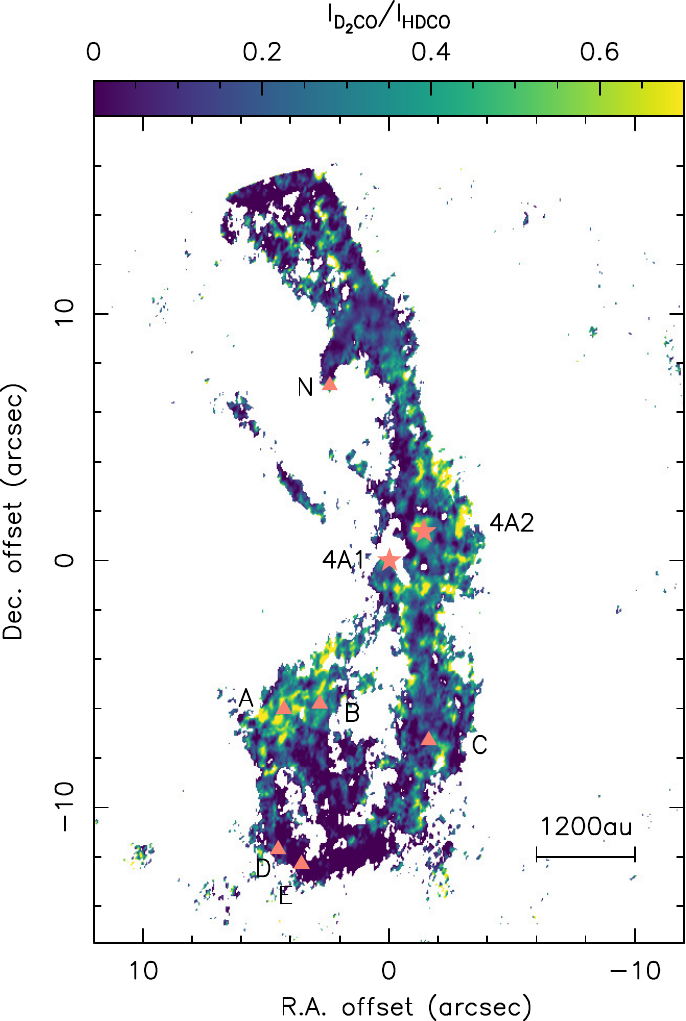}
    \caption{Intensity Ratio Map of D$_2$CO/HDCO toward the NGC\,1333 IRAS\,4A system. 
    The colour scale is shown at the top of the panel. 
    The two protostars are depicted with salmon stars. 
    The positions where the spectra are extracted are depicted with salmon triangles.}
    \label{fig:hdco-d2co-ratio}
\end{figure}

In Figure \ref{fig:hdco-d2co-ratio}, we present the intensity ratio map of D$_2$CO/HDCO, where we utilise the HDCO(4$_{1,4}$--3$_{1,3}$) transition integrated over a velocity interval (between 4 and 10\,km\,s$^{-1}$) corresponding to the D$_2$CO emission. While interpreting the region to the north of the protostar is intricate due to the overlapping emission from 4A1 and 4A2 cavities, we observe a notable gradient in the intensity ratio towards the southern region. Specifically, we observe a decrease in the D$_2$CO/HDCO ratio as we move towards the bow-shock, indicating a lower ratio at larger distances from 4A1. Additionally, we observe that the intensity ratio is lower in the western cavity wall compared to the eastern wall.

To ensure that this observed trend is not influenced by temperature variations along the cavity walls and the bow-shock, we estimate the gas temperature at five different locations along the southern cavity (A--E) marked by salmon triangles in the figure (the spectra are shown in Fig. \ref{fig:spectra}). Employing the local thermodynamic equilibrium (LTE) assumption and assuming optically thin lines, we conduct a rotational diagram (RD) analysis using HDCO lines (See App. \ref{appendix:lte} for details). Our analysis yields temperatures ranging from $\sim$23\,K to 27\,K with an uncertainty of $\pm$\,6\,K (see Table \ref{tab:deuteration-results}). Hence, we suggest that the disparity in the intensity ratio between the cavity walls and the bow-shock to the south of 4A1 is not attributable to temperature effects.

We calculate the deuteration ratio at the five marked positions using the intensities of the HDCO(4$_{1,4}$--3$_{1,3}$) and D$_2$CO lines, assuming an excitation temperature T$_{\rm{ex}}$ of 25.0$\pm$6.0 K.
To this end, we assume the LTE conditions and compute the [D$_{2}$]/[D] ratio from the HDCO(4$_{1,4}$--3$_{1,3}$) and D$_2$CO lines as follows:
\begin{equation}
  \begin{split}
        \rm{[{D_{2}]/[D]}} ~=~ & \frac{\rm{I_{D_2CO}}}{\rm{I_{HDCO}}}
        \times \frac{g_{\rm{HDCO}} ~\nu_{\rm{HDCO}} ~A_{\rm{HDCO}}}{g_{\rm{D_2CO}} ~\nu_{\rm{D_2CO}} ~A_{\rm{D_2CO}}}\\
        & \times \frac{Q_{\rm{D_2CO}}(\rm{T_{ex}})}{Q_{\rm{HDCO}}(\rm{T_{ex}})} 
        \times \exp\left[\frac{E_{\rm{D_2CO}}-E_{\rm{HDCO}}}{\rm{kT_{ex}}}\right]
  \end{split}
\end{equation}
where $g$, $\nu$, $A$ and $E$ are the upper-level statistical weight, frequency, spontaneous emission coefficient, and upper-level energy of the two used lines, and $Q(\rm{T_{ex}}$) is the partition function at $T_{\mathrm{ex}}$.

The derived values are summarised in Table \ref{tab:deuteration-results}. Specifically, we find that D$_{2}$/D ratio is approximately $\sim$45\% in the eastern cavity wall, decreasing to $\sim$25\% in the western cavity wall, and further reducing to $\sim$10\% towards the bow-shock, indicating a decrease in the deuteration ratio at larger distances from 4A1.

We note that performing a similar analysis at the north of 4A1 is challenging due to the overlap between the emission from 4A1 and 4A2 cavities. However, for consistency, we measured T$_{\rm{ex}}$ and derived the [D$_{2}$]/[D] ratio at the northern bow-shock (position N in Fig. \ref{fig:hdco-d2co-ratio}), where the emission is only due to the 4A1 outflow. We found slightly higher T$_{\rm{ex}}$ compared to the south, but they are consistent within the 1\,$\sigma$ uncertainty. Interestingly, the obtained deuteration ratio at this position is consistent with those obtained at the points (A--C) located at similar distances from 4A1 but to the south.

In addition, to verify the accuracy of the derived ratio at the bow-shock (located at the edge of the field of view), we estimated the deuteration ratio at the same position using the cubes imaged with a 200k$\lambda$ taper and found consistent values.


\section{Discussion \& Conclusions} \label{sec:discussion}

\begin{figure}
    \centering
    \includegraphics[width=0.46\textwidth]{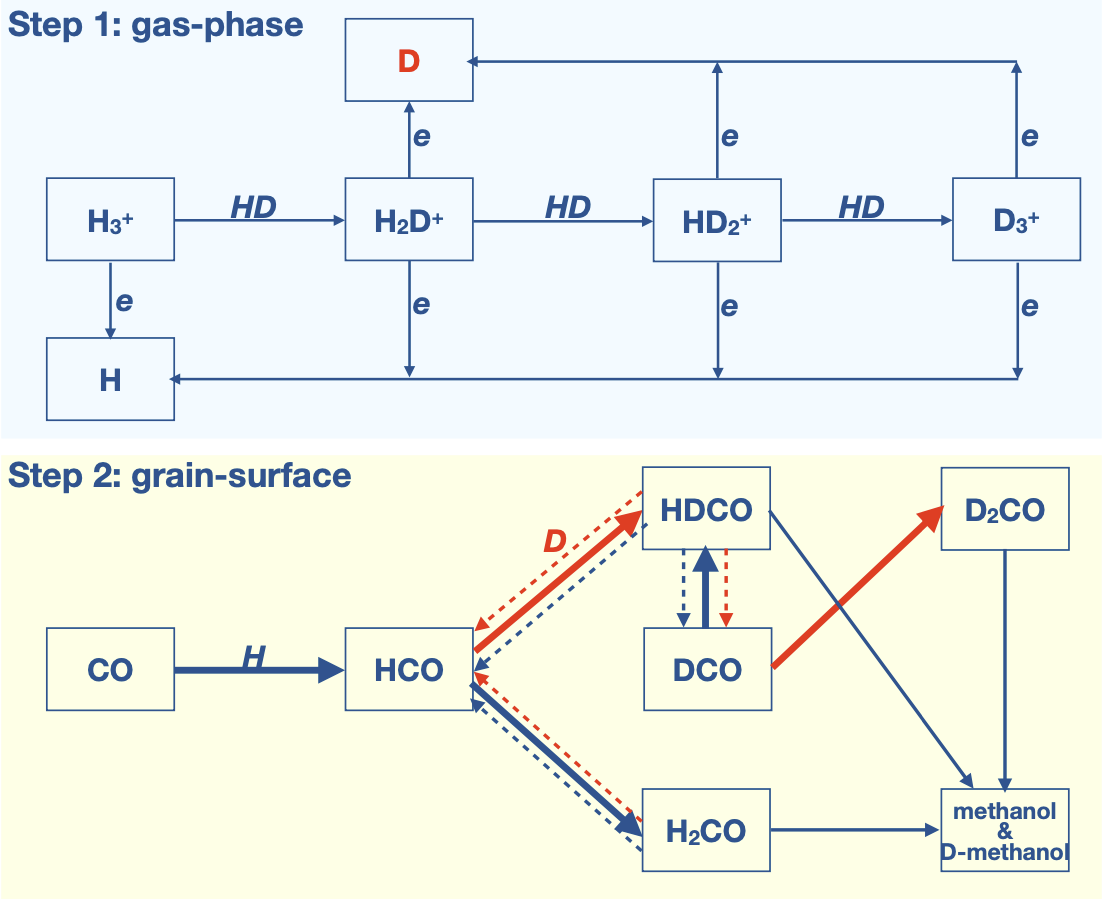}
    \caption{Scheme of the formation of HDCO and D$_2$CO from frozen CO on the grain surfaces, which occurs in two steps: 
    (1) the formation of atomic D (red) and H (blue) in the gas-phase (upper panel) and (2) the H (blue) and D (red) addition (solid lines) and abstraction (dashed lines) reactions to and from frozen CO on the grain-surfaces.
    The thick lines show the barrierless reactions, while the thin ones show the reactions with activation barriers \citep{Hidaka2009-D2COeDMethanol, Rimola2014, Song2017-H2CO&KIE} (see text).}
    \label{fig:D2CO-scheme}
\end{figure}


\noindent In principle, formaldehyde can form both in the gas-phase \citep[e.g.][]{Roueff2007-warmDeuteration} or on the grain-surfaces, by consecutive addition of H atoms \citep[e.g.][]{tielens1982}.
Its observed deuteration, in particular the measured [D$_2$CO]/[H$_2$CO] and [D$_2$CO]/[HDCO] ratios, helps in identifying the major route of H$_2$CO formation in an object where D$_2$CO is observed \citep[e.g.][]{Turner1990-D2CO, Ceccarelli1998-D2CO}.
Notably, large [D$_2$CO]/[HDCO] abundance ratios in lukewarm/warm ($\geq 20$ K) gas, similar to the ones measured in the present work along the IRAS\,4A1 cavities (0.1--0.6), exclude a major contribution of the gas-phase chemistry \cite[see, e.g., the theoretical predictions by][]{Bergman2011-d2co}.
Therefore, it is reasonable to assume that the HDCO and D$_2$CO detected in the IRAS\,4A cavities are grain-surface products.
Figure \ref{fig:D2CO-scheme} schematically recalls the two major steps at the base of their formation \citep[see, e.g., the reviews by][]{Ceccarelli2014-PP6, TsugeWatanabe2023-RevPJAB}.

\noindent
\textit{Step 1, gas-phase (upper panel):} 
D and H atoms are produced in the gas-phase by the electron recombination of the deuterated isotopologues of H$_3^+$.
Thus, the [D]/[H] atomic gas is directly proportional to the [H$_2$D$^+$]/[H$_3^+$] abundance ratio, which  is approximately given by:
\begin{equation}
 \frac{\rm{H_2 D^{+}}}{\rm{H_3^{+}}}=\frac{2 A_D k_{\text {form }}}{k_{\text {rec }} x_e+k_{CO} x_{CO}+2 k_{\text {form }} \exp [-E_{H_2D^+}/T]+2 k_{HD_2^+} A_D} 
\label{eq:deuteration} 
\end{equation}
where $A_D$ represents the elemental abundance of deuterium, $k_{\text{form}}$ is the rate constant for the formation of H$_2$D$^+$, $k_{\text{rec}}$ is the rate constant for the recombination of electrons with ions, $x_e$ is the electron abundance, $k_{\text{CO}}$ is the rate constant for the destruction of H$_2$D$^+$ due to reactions involving CO, $x_{\text{CO}}$ is the gaseous CO abundance, $E_{H_2D^+}$ is the exothermic energy for the formation reaction \citep[232 K:][]{Gerlich2002-H2D+energy}, $T$ is the gas kinetic temperature and $k_{HD_2^+}$ is the rate constant for the destruction of H$_2$D$^+$ due to reactions with HD forming HD$_2^+$.
Please note that Eq. \ref{eq:deuteration} only considers the singly deuterated isotopologue of H$_3^+$ to show the major parameters from which the [D]/[H] abundance ratio approximately depends.

\noindent
\textit{Step 2, grain-surface (lower panel):} 
Frozen CO is hydrogenated by consecutive addition of H and D atoms landing on the grain surfaces, leading to the H$_2$CO and CH$_3$OH formation, along with their isotopologues.
Two steps (CO + H and H$_2$CO + H) have large activation barriers, so they occur thanks to the H and D atoms tunnelling through them \citep[e.g.][]{Watanabe2002-methanol, Woon2002-COhydrogenation, Goumans2011-Dformaldehyde, Rimola2014, Song2017-H2CO&KIE}.
Given their larger mass, D tunnelling is much less efficient \citep[for example, at $\leq$50 K, the addition of D to H$_2$CO is about 100 times slower than that of H:][]{Song2017-H2CO&KIE}, so that D addition actually only occurs with radicals, e.g. HCO + D $\rightarrow$ HDCO.
In other words, D$_2$CO formation is not a simple sequence of D additions.
Experimental and theoretical works show that D and H abstraction reactions are in competition with the addition ones \citep{Hidaka2009-D2COeDMethanol, Goumans2011-Dformaldehyde, Song2017-H2CO&KIE} and are crucially important in the formation of the deuterated isotopologues of formaldehyde and methanol \citep{Taquet2012-Deuteration, Aikawa2012-Dchemistry, Riedel2023-Dmethanol}.

Nonetheless, D$_2$CO is formed via a sequence of D and H additions and abstractions, which lead to a [D$_2$CO]/[HDCO] governed by the gaseous [D]/[H] ratio.
The latter is, in turn, linked to the enhanced [H$_2$D$^+$]/[H$_3^+$] abundance ratio to the first order \citep[the other isotopologues of H$_3^+$ also contribute to D/H: e.g.,][]{Vastel2004-D2H+}.
Equation \ref{eq:deuteration} indicates that three major parameters enter into play: the gas ionisation and temperature, and the gaseous CO (and heavy species) abundance.
The lower each of them is, the higher the [H$_2$D$^+$]/[H$_3^+$] abundance ratio and, consequently, [D]/[H] and [D$_2$CO]/[HDCO].
Notably, for a given cosmic-ray ionization rate, the electronic abundance is inversely proportional to the square of density\footnote{Assuming that the ionization is dominated by the cosmic-rays and the major positive charge carrier is H$_3^+$ and its isotopologues, it is to demonstrate that the electron density is proportional to the inverse of the square of the density \citep[e.g.,][]{McKee1989-ionization}.}. 
Likewise,  gaseous CO (and heavy species) abundance also roughly diminishes with increasing gas density, so that the [D$_2$CO]/[HDCO] abundance ratio mainly depends on two parameters: the gas density and temperature.
Density is high and temperature extremely low in prestellar cores and, for this reason, grain mantles enriched with deuterated formaldehyde are thought to be mostly formed during this phase \citep[e.g.][]{Turner1990-D2CO, Ceccarelli1998-D2CO, Taquet2012-Deuteration}.

Our analysis reveals a consistent decrease in the [D$_2$CO]/[HDCO] ratio at larger distances from the protostar along the outflow cavity of IRAS\,4A1 (Tab. \ref{tab:deuteration-results}). 
This variation could stem from two distinct possibilities: 
it may be entirely inherited from the prestellar core phase, reflecting the pristine ice composition, or it could have been altered later by gas-phase reactions, leading to diminished deuteration levels, especially in regions characterised by relatively high gas temperatures, such as shocked gas environments. 
The latter scenario typically becomes prominent approximately $10^5$ years after the injection of HDCO and D$_2$CO into the gas phase \citep[see, e.g., Fig. 4 in][]{Charnley1997-Dmethanol}. 
Considering that the IRAS\,4A system is younger than $10^5$ years and, therefore, its outflow cavities are also younger than this age, we suggest that this pathway may not yet be significantly active in our case. 
Subsequently, we concentrate on exploring the implications of the first scenario, namely that the observed [D$_2$CO]/[HDCO] ratio is dominated by the conditions in the prestellar core precursor of IRAS\,4A.

Our observations show that, between $\sim$\,1900 and 2300 au, the [D$_2$CO]/[HDCO] ratio shows minimal variation within uncertainties (approximately $\sim\,41.5^{+8.9}_{-7.9}$\% at $\sim$\,1900\,au and $\sim\,30.3^{+8.5}_{-6.5}$\% at $\sim$2300 au), aligning within uncertainties with values reported by \citet[][$\sim\,62.0^{+33.0}_{-26.0}$\%]{Parise2006} within $\sim$\,3000\,au. 
This suggests that the prestellar core preceding the IRAS\,4A protostellar system had a relatively flat density profile at distances < 3000\,au. 
In addition, the notable decrease to $\sim$\,10\% in the ratio towards the bow-shock (at $\sim$4000\,au) implies a significant decrease in the density profile (of the  IRAS\,4A prestellar core) beyond 3000\,au. 
Consequently, we propose that while the  IRAS\,4A prestellar core had a roughly constant density profile within the inner $\sim$\,3000\,au, a substantial decrease in density occurred beyond this radius, leading to the observed decline in the deuteration ratio.
This is in agreement with the expected and measured density profile of prestellar cores, where the density drastically decreases towards the outer regions and increases towards the inner regions, reaching a plateau \citep[e.g.,][]{Foster1993-PSCstructure, Evans2001-PSCstructure, Keto2010, Nielbock2012-B68structure, Lippok2016-PSCstructure}. 
Remarkably, the extreme deuteration observed within a radius of about 3000 au of the prestellar core preceding IRAS\,4A coincides with the almost complete freeze-out zone of the prestellar core L1544, where the density profile has indeed a plateau and the highest deuteration of species formed by hydrogenation on the grain surfaces is expected \citep{Caselli2022-centerL1544-freeze}.

Additionally, considering the extremities of the obtained values, the deuteration ratio at a distance of 2000 au ranges from 20-60\%, decreasing to 10\% at 4000\,au. 
This implies a decrease in deuteration by a factor of 2-6 from 2000 to 4000\,au. 
Assuming that the CO abundance of the prestellar core is low enough so that the electron recombination is the major H$_2$D$^+$ destruction route (Eq. \ref{eq:deuteration}) and the [D$_2$CO]/[HDCO] variation is solely due to the density decrease (with [D]/[H] $\propto\,x_{e}^{-1}\,\propto\,n_{\rm{H_{2}}}^{1/2}$
), it would correspond to a density variation of 4-36. 
However, in the outer regions of a prestellar core, density variation roughly follows a $r^{-2}$ dependence (see above references), suggesting a density decrease by a factor of 4 between 2000 and 4000\,au. 
Therefore, unless the prestellar core of IRAS\,4A had a steeper outer density profile \citep[as the case of B68;][]{Nielbock2012-B68structure}, the observed trend in deuteration may also indicate a decreasing temperature profile from the outer to the inner regions, as observed in present prestellar cores \citep[e.g.,][]{Crapsi2007-PSCtemp, Keto2010, Lippok2016-PSCstructure}.

In conclusion, in this Letter we present a new method for studying the original prestellar core structure of current protostars by observing species injected by outflow-triggered shocks near the protostar's centre.

Deuterated species, such as the formaldehyde D-isotopologues, are particularly powerful because their relative abundances depend on the density and temperature profiles of the prestellar core.
Notice that, in the innermost regions of prestellar cores, the molecular deuteration profile is inaccessible because all species freeze out onto the grain mantles, except H and D only bearing species.
Therefore, the proposed method is also the only one able to provide molecular deuteration in those crucial regions.

\section*{Acknowledgments}
This article makes use of the following ALMA data: ADS/JAO.ALMA2018.1.01205.L (PI: S. Yamamoto). 
ALMA is a partnership of the ESO (representing its member states), the NSF (USA) and NINS (Japan), together with the NRC (Canada) and the NSC and ASIAA (Taiwan), in cooperation with the Republic of Chile. The Joint ALMA Observatory is operated by the ESO, the AUI/NRAO, and the NAOJ. 
The authors thank the ALMA and NRAO staff for their support. 
Part of the data reduction/combination presented in this paper was performed using the GRICAD infrastructure (\url{https://gricad.univ-grenoble-alpes.fr}).
This project has received funding from the European Research Council (ERC) under the European Union's Horizon 2020 research and innovation program, for the Project “The Dawn of Organic Chemistry” (DOC), grant agreement No 741002 and the European Union’s Horizon 2020 research and innovation programs under projects “Astro-Chemistry Origins” (ACO), Grant No 811312.
ClCo, LP, GS and EB acknowledge the PRIN-MUR 2020 BEYOND-2p (``Astrochemistry beyond the second period elements'', Prot. 2020AFB3FX), the project ASI-Astrobiologia 2023 MIGLIORA (Modeling Chemical Complexity, F83C23000800005), the INAF-GO 2023 fundings PROTO-SKA (Exploiting ALMA data to study planet forming disks: preparing the advent of SKA, C13C23000770005), the INAF Mini-Grant 2022 “Chemical Origins” (PI: L. Podio), INAF-Minigrant 2023 TRIESTE (``TRacing the chemIcal hEritage of our originS: from proTostars to planEts''; PI: G. Sabatini), and the National Recovery and Resilience Plan (NRRP), Mission 4, Component 2, Investment 1.1, Call for tender No. 104 published on 2.2.2022 by the Italian Ministry of University and Research (MUR), funded by the European Union – NextGenerationEU– Project Title 2022JC2Y93 Chemical Origins: linking the fossil composition of the Solar System with the chemistry of protoplanetary disks – CUP J53D23001600006 - Grant Assignment Decree No. 962 adopted on 30.06.2023 by the Italian Ministry of University and Research (MUR).
LL acknowledges the support of UNAM-DGAPA PAPIIT grants IN108324 and IN112820 and CONACYT-CF grant 263356.
EB acknowledges support from the Deutsche Forschungsgemeinschaft (DFG, German Research Foundation) under German´s Excellence Strategy – EXC 2094 – 390783311.
MB acknowledges support from the European Research Council (ERC) Advanced Grant MOPPEX 833460.
SBC was supported by the NASA Planetary Science Division Internal Scientist Funding Pro- gram through the Fundamental Laboratory Research work package (FLaRe).
NC acknowledges support from the European Research Council (ERC) project ‘Stellar-MADE’ (No.101042275). 
This study is supported by the MEXT/ JSPS Grant-in-Aid from the Ministry of Education, Culture, Sports, Science, and Technology of Japan (JP20H05844 and JP20H05845). 

\section*{Data Availability}
The raw data are available on the ALMA archive at the end of the proprietary period (ADS/JAO.ALMA\#2018.1.01205.L).


\bibliographystyle{mnras}
\bibliography{CeciliaCeccarelli} 

\begin{thebibliography}{}
\makeatletter
\relax
\def\mn@urlcharsother{\let\do\@makeother \do\$\do\&\do\#\do\^\do\_\do\%\do\~}
\def\mn@doi{\begingroup\mn@urlcharsother \@ifnextchar [ {\mn@doi@} {\mn@doi@[]}}
\def\mn@doi@[#1]#2{\def\@tempa{#1}\ifx\@tempa\@empty \href {http://dx.doi.org/#2} {doi:#2}\else \href {http://dx.doi.org/#2} {#1}\fi \endgroup}
\def\mn@eprint#1#2{\mn@eprint@#1:#2::\@nil}
\def\mn@eprint@arXiv#1{\href {http://arxiv.org/abs/#1} {{\tt arXiv:#1}}}
\def\mn@eprint@dblp#1{\href {http://dblp.uni-trier.de/rec/bibtex/#1.xml} {dblp:#1}}
\def\mn@eprint@#1:#2:#3:#4\@nil{\def\@tempa {#1}\def\@tempb {#2}\def\@tempc {#3}\ifx \@tempc \@empty \let \@tempc \@tempb \let \@tempb \@tempa \fi \ifx \@tempb \@empty \def\@tempb {arXiv}\fi \@ifundefined {mn@eprint@\@tempb}{\@tempb:\@tempc}{\expandafter \expandafter \csname mn@eprint@\@tempb\endcsname \expandafter{\@tempc}}}

\bibitem[\protect\citeauthoryear{{Aikawa}, {Wakelam}, {Hersant}, {Garrod}  \& {Herbst}}{{Aikawa} et~al.}{2012}]{Aikawa2012-Dchemistry}
{Aikawa} Y.,  {Wakelam} V.,  {Hersant} F.,  {Garrod} R.~T.,   {Herbst} E.,  2012, \mn@doi [\apj] {10.1088/0004-637X/760/1/40}, \href {https://ui.adsabs.harvard.edu/abs/2012ApJ...760...40A} {760, 40}

\bibitem[\protect\citeauthoryear{{Bacmann}, {Lefloch}, {Ceccarelli}, {Castets}, {Steinacker}  \& {Loinard}}{{Bacmann} et~al.}{2002}]{Bacmann2002-COdepletion}
{Bacmann} A.,  {Lefloch} B.,  {Ceccarelli} C.,  {Castets} A.,  {Steinacker} J.,   {Loinard} L.,  2002, \mn@doi [\aap] {10.1051/0004-6361:20020652}, \href {https://ui.adsabs.harvard.edu/abs/2002A&A...389L...6B} {389, L6}

\bibitem[\protect\citeauthoryear{{Bacmann}, {Lefloch}, {Ceccarelli}, {Steinacker}, {Castets}  \& {Loinard}}{{Bacmann} et~al.}{2003}]{Bacmann2003-COdepD2CO}
{Bacmann} A.,  {Lefloch} B.,  {Ceccarelli} C.,  {Steinacker} J.,  {Castets} A.,   {Loinard} L.,  2003, \mn@doi [\apjl] {10.1086/374263}, \href {http://adsabs.harvard.edu/abs/2003ApJ...585L..55B} {585, L55}

\bibitem[\protect\citeauthoryear{{Bergman}, {Parise}, {Liseau}  \& {Larsson}}{{Bergman} et~al.}{2011}]{Bergman2011-d2co}
{Bergman} P.,  {Parise} B.,  {Liseau} R.,   {Larsson} B.,  2011, \mn@doi [\aap] {10.1051/0004-6361/201015012}, \href {https://ui.adsabs.harvard.edu/abs/2011A&A...527A..39B} {527, A39}

\bibitem[\protect\citeauthoryear{{Bianchi} et~al.,}{{Bianchi} et~al.}{2017}]{Bianchi2017-SVS13deuteration}
{Bianchi} E.,  et~al., 2017, \mn@doi [\mnras] {10.1093/mnras/stx252}, \href {http://adsabs.harvard.edu/abs/2017MNRAS.467.3011B} {467, 3011}

\bibitem[\protect\citeauthoryear{{Blake}, {Sandell}, {van Dishoeck}, {Groesbeck}, {Mundy}  \& {Aspin}}{{Blake} et~al.}{1995}]{Blake1995}
{Blake} G.~A.,  {Sandell} G.,  {van Dishoeck} E.~F.,  {Groesbeck} T.~D.,  {Mundy} L.~G.,   {Aspin} C.,  1995, \mn@doi [\apj] {10.1086/175392}, \href {https://ui.adsabs.harvard.edu/abs/1995ApJ...441..689B} {441, 689}

\bibitem[\protect\citeauthoryear{{Bocquet} et~al.,}{{Bocquet} et~al.}{1999}]{Bocquet1999}
{Bocquet} R.,  et~al., 1999, \mn@doi [Journal of Molecular Spectroscopy] {10.1006/jmsp.1999.7836}, \href {https://ui.adsabs.harvard.edu/abs/1999JMoSp.195..345B} {195, 345}

\bibitem[\protect\citeauthoryear{{CASA Team} et~al.,}{{CASA Team} et~al.}{2022}]{CasaTeam2022}
{CASA Team} et~al., 2022, \mn@doi [\pasp] {10.1088/1538-3873/ac9642}, \href {https://ui.adsabs.harvard.edu/abs/2022PASP..134k4501C} {134, 114501}

\bibitem[\protect\citeauthoryear{{Caselli}, {Walmsley}, {Tafalla}, {Dore}  \& {Myers}}{{Caselli} et~al.}{1999}]{Caselli1999-COdepl}
{Caselli} P.,  {Walmsley} C.~M.,  {Tafalla} M.,  {Dore} L.,   {Myers} P.~C.,  1999, \mn@doi [\apjl] {10.1086/312280}, \href {http://adsabs.harvard.edu/abs/1999ApJ...523L.165C} {523, L165}

\bibitem[\protect\citeauthoryear{{Caselli}, {van der Tak}, {Ceccarelli}  \& {Bacmann}}{{Caselli} et~al.}{2003}]{Caselli2003-h2dp}
{Caselli} P.,  {van der Tak} F.~F.~S.,  {Ceccarelli} C.,   {Bacmann} A.,  2003, \mn@doi [\aap] {10.1051/0004-6361:20030526}, \href {https://ui.adsabs.harvard.edu/abs/2003A&A...403L..37C} {403, L37}

\bibitem[\protect\citeauthoryear{{Caselli} et~al.,}{{Caselli} et~al.}{2022}]{Caselli2022-centerL1544-freeze}
{Caselli} P.,  et~al., 2022, \mn@doi [\apj] {10.3847/1538-4357/ac5913}, \href {https://ui.adsabs.harvard.edu/abs/2022ApJ...929...13C} {929, 13}

\bibitem[\protect\citeauthoryear{{Ceccarelli}, {Castets}, {Loinard}, {Caux}  \& {Tielens}}{{Ceccarelli} et~al.}{1998}]{Ceccarelli1998-D2CO}
{Ceccarelli} C.,  {Castets} A.,  {Loinard} L.,  {Caux} E.,   {Tielens} A.~G.~G.~M.,  1998, \aap, \href {http://adsabs.harvard.edu/abs/1998A%26A...338L..43C} {338, L43}

\bibitem[\protect\citeauthoryear{{Ceccarelli}, {Loinard}, {Castets}, {Tielens}, {Caux}, {Lefloch}  \& {Vastel}}{{Ceccarelli} et~al.}{2001}]{Ceccarelli2001-extendedD2CO}
{Ceccarelli} C.,  {Loinard} L.,  {Castets} A.,  {Tielens} A.~G.~G.~M.,  {Caux} E.,  {Lefloch} B.,   {Vastel} C.,  2001, \mn@doi [\aap] {10.1051/0004-6361:20010559}, \href {http://adsabs.harvard.edu/abs/2001A%26A...372..998C} {372, 998}

\bibitem[\protect\citeauthoryear{{Ceccarelli}, {Vastel}, {Tielens}, {Castets}, {Boogert}, {Loinard}  \& {Caux}}{{Ceccarelli} et~al.}{2002}]{Ceccarelli2002-D2COinL1689N}
{Ceccarelli} C.,  {Vastel} C.,  {Tielens} A.~G.~G.~M.,  {Castets} A.,  {Boogert} A.~C.~A.,  {Loinard} L.,   {Caux} E.,  2002, \mn@doi [\aap] {10.1051/0004-6361:20011547}, \href {https://ui.adsabs.harvard.edu/abs/2002A&A...381L..17C} {381, L17}

\bibitem[\protect\citeauthoryear{{Ceccarelli}, {Caselli}, {Bockel{\'e}e-Morvan}, {Mousis}, {Pizzarello}, {Robert}  \& {Semenov}}{{Ceccarelli} et~al.}{2014}]{Ceccarelli2014-PP6}
{Ceccarelli} C.,  {Caselli} P.,  {Bockel{\'e}e-Morvan} D.,  {Mousis} O.,  {Pizzarello} S.,  {Robert} F.,   {Semenov} D.,  2014, \mn@doi [Protostars and Planets VI] {10.2458/azu_uapress_9780816531240-ch037}, \href {http://adsabs.harvard.edu/abs/2014prpl.conf..859C} {pp 859--882}

\bibitem[\protect\citeauthoryear{{Chac{\'o}n-Tanarro} et~al.,}{{Chac{\'o}n-Tanarro} et~al.}{2019}]{Chacon2019-Dmethanol}
{Chac{\'o}n-Tanarro} A.,  et~al., 2019, \mn@doi [\aap] {10.1051/0004-6361/201832703}, \href {http://adsabs.harvard.edu/abs/2019AM26A...622A.141C} {622, A141}

\bibitem[\protect\citeauthoryear{{Chahine} et~al.,}{{Chahine} et~al.}{2024}]{Chahine2024}
{Chahine} L.,  et~al., 2024, \mn@doi [\mnras] {10.1093/mnras/stae1320}, \href {https://ui.adsabs.harvard.edu/abs/2024MNRAS.531.2653C} {531, 2653}

\bibitem[\protect\citeauthoryear{{Charnley}, {Tielens}  \& {Rodgers}}{{Charnley} et~al.}{1997}]{Charnley1997-Dmethanol}
{Charnley} S.~B.,  {Tielens} A.~G.~G.~M.,   {Rodgers} S.~D.,  1997, \mn@doi [\apjl] {10.1086/310697}, \href {http://adsabs.harvard.edu/abs/1997ApJ...482L.203C} {482, L203}

\bibitem[\protect\citeauthoryear{{Ching}, {Lai}, {Zhang}, {Yang}, {Girart}  \& {Rao}}{{Ching} et~al.}{2016}]{Ching2016}
{Ching} T.-C.,  {Lai} S.-P.,  {Zhang} Q.,  {Yang} L.,  {Girart} J.~M.,   {Rao} R.,  2016, \mn@doi [\apj] {10.3847/0004-637X/819/2/159}, \href {https://ui.adsabs.harvard.edu/abs/2016ApJ...819..159C} {819, 159}

\bibitem[\protect\citeauthoryear{{Choi}}{{Choi}}{2001}]{Choi2001}
{Choi} M.,  2001, \mn@doi [\apj] {10.1086/320657}, \href {https://ui.adsabs.harvard.edu/abs/2001ApJ...553..219C} {553, 219}

\bibitem[\protect\citeauthoryear{{Choi}}{{Choi}}{2005}]{Choi2005}
{Choi} M.,  2005, \mn@doi [\apj] {10.1086/432113}, \href {https://ui.adsabs.harvard.edu/abs/2005ApJ...630..976C} {630, 976}

\bibitem[\protect\citeauthoryear{{Chuang}, {Aso}, {Hirano}, {Hirano}  \& {Machida}}{{Chuang} et~al.}{2021}]{Chuang2021}
{Chuang} C.-Y.,  {Aso} Y.,  {Hirano} N.,  {Hirano} S.,   {Machida} M.~N.,  2021, \mn@doi [\apj] {10.3847/1538-4357/abfdbb}, \href {https://ui.adsabs.harvard.edu/abs/2021ApJ...916...82C} {916, 82}

\bibitem[\protect\citeauthoryear{{Codella}, {Ceccarelli}, {Chandler}, {Sakai}, {Yamamoto}  \& {FAUST Team}}{{Codella} et~al.}{2021}]{Codella2021}
{Codella} C.,  {Ceccarelli} C.,  {Chandler} C.,  {Sakai} N.,  {Yamamoto} S.,   {FAUST Team} 2021, \mn@doi [Fr.Astr.Sp.Sci.] {10.3389/fspas.2021.782006}, \href {https://ui.adsabs.harvard.edu/abs/2021FrASS...8..227C} {8, 227}

\bibitem[\protect\citeauthoryear{{Cooke}, {Pettini}  \& {Steidel}}{{Cooke} et~al.}{2018}]{Cooke2018-cosmicD}
{Cooke} R.~J.,  {Pettini} M.,   {Steidel} C.~C.,  2018, \mn@doi [\apj] {10.3847/1538-4357/aaab53}, \href {https://ui.adsabs.harvard.edu/abs/2018ApJ...855..102C} {855, 102}

\bibitem[\protect\citeauthoryear{{Coutens} et~al.,}{{Coutens} et~al.}{2013}]{Coutens2013}
{Coutens} A.,  et~al., 2013, \mn@doi [\aap] {10.1051/0004-6361/201322400}, \href {https://ui.adsabs.harvard.edu/abs/2013A&A...560A..39C} {560, A39}

\bibitem[\protect\citeauthoryear{{Crapsi}, {Caselli}, {Walmsley}, {Myers}, {Tafalla}, {Lee}  \& {Bourke}}{{Crapsi} et~al.}{2005}]{Crapsi2005}
{Crapsi} A.,  {Caselli} P.,  {Walmsley} C.~M.,  {Myers} P.~C.,  {Tafalla} M.,  {Lee} C.~W.,   {Bourke} T.~L.,  2005, \mn@doi [\apj] {10.1086/426472}, \href {https://ui.adsabs.harvard.edu/abs/2005ApJ...619..379C} {619, 379}

\bibitem[\protect\citeauthoryear{{Crapsi}, {Caselli}, {Walmsley}  \& {Tafalla}}{{Crapsi} et~al.}{2007}]{Crapsi2007-PSCtemp}
{Crapsi} A.,  {Caselli} P.,  {Walmsley} M.~C.,   {Tafalla} M.,  2007, \mn@doi [\aap] {10.1051/0004-6361:20077613}, \href {https://ui.adsabs.harvard.edu/abs/2007A&A...470..221C} {470, 221}

\bibitem[\protect\citeauthoryear{{De Simone} et~al.,}{{De Simone} et~al.}{2020}]{Desimone2020-I4outflow}
{De Simone} M.,  et~al., 2020, \mn@doi [\aap] {10.1051/0004-6361/201937004}, \href {https://ui.adsabs.harvard.edu/abs/2020A&A...640A..75D} {640, A75}

\bibitem[\protect\citeauthoryear{{Di Francesco}, {Myers}, {Wilner}, {Ohashi}  \& {Mardones}}{{Di Francesco} et~al.}{2001}]{DiFrancesco2001}
{Di Francesco} J.,  {Myers} P.~C.,  {Wilner} D.~J.,  {Ohashi} N.,   {Mardones} D.,  2001, \mn@doi [\apj] {10.1086/323854}, \href {https://ui.adsabs.harvard.edu/abs/2001ApJ...562..770D} {562, 770}

\bibitem[\protect\citeauthoryear{{Evans}, {Rawlings}, {Shirley}  \& {Mundy}}{{Evans} et~al.}{2001}]{Evans2001-PSCstructure}
{Evans} Neal~J. I.,  {Rawlings} J. M.~C.,  {Shirley} Y.~L.,   {Mundy} L.~G.,  2001, \mn@doi [\apj] {10.1086/321639}, \href {https://ui.adsabs.harvard.edu/abs/2001ApJ...557..193E} {557, 193}

\bibitem[\protect\citeauthoryear{{Evans} et~al.,}{{Evans} et~al.}{2023}]{Evans2023}
{Evans} L.,  et~al., 2023, \mn@doi [\aap] {10.1051/0004-6361/202346428}, \href {https://ui.adsabs.harvard.edu/abs/2023A&A...678A.160E} {678, A160}

\bibitem[\protect\citeauthoryear{{Fontani} et~al.,}{{Fontani} et~al.}{2011}]{Fontani2011-highmassDeut}
{Fontani} F.,  et~al., 2011, \mn@doi [\aap] {10.1051/0004-6361/201116631}, \href {https://ui.adsabs.harvard.edu/abs/2011A&A...529L...7F} {529, L7}

\bibitem[\protect\citeauthoryear{{Foster} \& {Chevalier}}{{Foster} \& {Chevalier}}{1993}]{Foster1993-PSCstructure}
{Foster} P.~N.,  {Chevalier} R.~A.,  1993, \mn@doi [\apj] {10.1086/173236}, \href {https://ui.adsabs.harvard.edu/abs/1993ApJ...416..303F} {416, 303}

\bibitem[\protect\citeauthoryear{{Gerlich}, {Herbst}  \& {Roueff}}{{Gerlich} et~al.}{2002}]{Gerlich2002-H2D+energy}
{Gerlich} D.,  {Herbst} E.,   {Roueff} E.,  2002, \mn@doi [\planss] {10.1016/S0032-0633(02)00094-6}, \href {https://ui.adsabs.harvard.edu/abs/2002P&SS...50.1275G} {50, 1275}

\bibitem[\protect\citeauthoryear{{Girart}, {Crutcher}  \& {Rao}}{{Girart} et~al.}{1999}]{Girart1999}
{Girart} J.~M.,  {Crutcher} R.~M.,   {Rao} R.,  1999, \mn@doi [\apjl] {10.1086/312345}, \href {https://ui.adsabs.harvard.edu/abs/1999ApJ...525L.109G} {525, L109}

\bibitem[\protect\citeauthoryear{{Goumans}}{{Goumans}}{2011}]{Goumans2011-Dformaldehyde}
{Goumans} T.~P.~M.,  2011, \mn@doi [\mnras] {10.1111/j.1365-2966.2011.18329.x}, \href {https://ui.adsabs.harvard.edu/abs/2011MNRAS.413.2615G} {413, 2615}

\bibitem[\protect\citeauthoryear{{Hatchell}}{{Hatchell}}{2003}]{Hatchell2003}
{Hatchell} J.,  2003, \mn@doi [\aap] {10.1051/0004-6361:20030297}, \href {https://ui.adsabs.harvard.edu/abs/2003A&A...403L..25H} {403, L25}

\bibitem[\protect\citeauthoryear{{Hidaka}, {Watanabe}, {Kouchi}  \& {Watanabe}}{{Hidaka} et~al.}{2009}]{Hidaka2009-D2COeDMethanol}
{Hidaka} H.,  {Watanabe} M.,  {Kouchi} A.,   {Watanabe} N.,  2009, \mn@doi [\apj] {10.1088/0004-637X/702/1/291}, \href {https://ui.adsabs.harvard.edu/abs/2009ApJ...702..291H} {702, 291}

\bibitem[\protect\citeauthoryear{{H{\"o}gbom}}{{H{\"o}gbom}}{1974}]{Hogbom1974}
{H{\"o}gbom} J.~A.,  1974, \aaps, \href {https://ui.adsabs.harvard.edu/abs/1974A&AS...15..417H} {15, 417}

\bibitem[\protect\citeauthoryear{{J{\o}rgensen} et~al.,}{{J{\o}rgensen} et~al.}{2016}]{Jorgensen2016}
{J{\o}rgensen} J.~K.,  et~al., 2016, \mn@doi [\aap] {10.1051/0004-6361/201628648}, \href {http://adsabs.harvard.edu/abs/2016A%26A...595A.117J} {595, A117}

\bibitem[\protect\citeauthoryear{{J{\o}rgensen} et~al.,}{{J{\o}rgensen} et~al.}{2018}]{Jorgensen2018}
{J{\o}rgensen} J.~K.,  et~al., 2018, \mn@doi [\aap] {10.1051/0004-6361/201731667}, \href {http://adsabs.harvard.edu/abs/2018A%26A...620A.170J} {620, A170}

\bibitem[\protect\citeauthoryear{{Keto} \& {Caselli}}{{Keto} \& {Caselli}}{2010}]{Keto2010}
{Keto} E.,  {Caselli} P.,  2010, \mn@doi [\mnras] {10.1111/j.1365-2966.2009.16033.x}, \href {https://ui.adsabs.harvard.edu/abs/2010MNRAS.402.1625K} {402, 1625}

\bibitem[\protect\citeauthoryear{{Koumpia}, {Semenov}, {van der Tak}, {Boogert}  \& {Caux}}{{Koumpia} et~al.}{2017}]{Koumpia2017}
{Koumpia} E.,  {Semenov} D.~A.,  {van der Tak} F.~F.~S.,  {Boogert} A.~C.~A.,   {Caux} E.,  2017, \mn@doi [\aap] {10.1051/0004-6361/201630160}, \href {https://ui.adsabs.harvard.edu/abs/2017A&A...603A..88K} {603, A88}

\bibitem[\protect\citeauthoryear{{Kristensen} et~al.,}{{Kristensen} et~al.}{2012}]{Kristensen2012}
{Kristensen} L.~E.,  et~al., 2012, \mn@doi [\aap] {10.1051/0004-6361/201118146}, \href {https://ui.adsabs.harvard.edu/abs/2012A&A...542A...8K} {542, A8}

\bibitem[\protect\citeauthoryear{{Lefloch}, {Castets}, {Cernicharo}, {Langer}  \& {Zylka}}{{Lefloch} et~al.}{1998}]{Lefloch1998a}
{Lefloch} B.,  {Castets} A.,  {Cernicharo} J.,  {Langer} W.~D.,   {Zylka} R.,  1998, \aap, \href {http://adsabs.harvard.edu/abs/1998A%26A...334..269L} {334, 269}

\bibitem[\protect\citeauthoryear{{Lippok} et~al.,}{{Lippok} et~al.}{2016}]{Lippok2016-PSCstructure}
{Lippok} N.,  et~al., 2016, \mn@doi [\aap] {10.1051/0004-6361/201525792}, \href {https://ui.adsabs.harvard.edu/abs/2016A&A...592A..61L} {592, A61}

\bibitem[\protect\citeauthoryear{{Liu}, {Parise}, {Kristensen}, {Visser}, {van Dishoeck}  \& {G{\"u}sten}}{{Liu} et~al.}{2011}]{Liu2011}
{Liu} F.~C.,  {Parise} B.,  {Kristensen} L.,  {Visser} R.,  {van Dishoeck} E.~F.,   {G{\"u}sten} R.,  2011, \mn@doi [\aap] {10.1051/0004-6361/201015519}, \href {https://ui.adsabs.harvard.edu/abs/2011A&A...527A..19L} {527, A19}

\bibitem[\protect\citeauthoryear{{Loinard}, {Castets}, {Ceccarelli}, {Caux}  \& {Tielens}}{{Loinard} et~al.}{2001}]{Loinard2001}
{Loinard} L.,  {Castets} A.,  {Ceccarelli} C.,  {Caux} E.,   {Tielens} A.~G.~G.~M.,  2001, \mn@doi [\apjl] {10.1086/320331}, \href {http://adsabs.harvard.edu/abs/2001ApJ...552L.163L} {552, L163}

\bibitem[\protect\citeauthoryear{{Loinard} et~al.,}{{Loinard} et~al.}{2002}]{Loinard2002}
{Loinard} L.,  et~al., 2002, \mn@doi [\planss] {10.1016/S0032-0633(02)00084-3}, \href {http://adsabs.harvard.edu/abs/2002P%26SS...50.1205L} {50, 1205}

\bibitem[\protect\citeauthoryear{{Manigand} et~al.,}{{Manigand} et~al.}{2020}]{Manigand2020-i16293a}
{Manigand} S.,  et~al., 2020, \mn@doi [\aap] {10.1051/0004-6361/201936299}, \href {https://ui.adsabs.harvard.edu/abs/2020A&A...635A..48M} {635, A48}

\bibitem[\protect\citeauthoryear{{Maury} et~al.,}{{Maury} et~al.}{2019}]{Maury2019}
{Maury} A.~J.,  et~al., 2019, \mn@doi [\aap] {10.1051/0004-6361/201833537}, \href {https://ui.adsabs.harvard.edu/abs/2019A&A...621A..76M} {621, A76}

\bibitem[\protect\citeauthoryear{{McKee}}{{McKee}}{1989}]{McKee1989-ionization}
{McKee} C.~F.,  1989, \mn@doi [\apj] {10.1086/167950}, \href {https://ui.adsabs.harvard.edu/abs/1989ApJ...345..782M} {345, 782}

\bibitem[\protect\citeauthoryear{{Mercimek} et~al.,}{{Mercimek} et~al.}{2022}]{Mercimek2022}
{Mercimek} S.,  et~al., 2022, \mn@doi [\aap] {10.1051/0004-6361/202141790}, \href {https://ui.adsabs.harvard.edu/abs/2022A&A...659A..67M} {659, A67}

\bibitem[\protect\citeauthoryear{{Nielbock} et~al.,}{{Nielbock} et~al.}{2012}]{Nielbock2012-B68structure}
{Nielbock} M.,  et~al., 2012, \mn@doi [\aap] {10.1051/0004-6361/201219139}, \href {https://ui.adsabs.harvard.edu/abs/2012A&A...547A..11N} {547, A11}

\bibitem[\protect\citeauthoryear{{Nomura} et~al.,}{{Nomura} et~al.}{2023}]{Nomura2023-PP7}
{Nomura} H.,  et~al., 2023, in {Inutsuka} S.,  {Aikawa} Y.,  {Muto} T.,  {Tomida} K.,   {Tamura} M.,  eds,  Astronomical Society of the Pacific Conference Series Vol. 534, Protostars and Planets VII. p.~1075

\bibitem[\protect\citeauthoryear{{Parise}, {Ceccarelli}, {Tielens}, {Castets}, {Caux}, {Lefloch}  \& {Maret}}{{Parise} et~al.}{2006}]{Parise2006}
{Parise} B.,  {Ceccarelli} C.,  {Tielens} A.~G.~G.~M.,  {Castets} A.,  {Caux} E.,  {Lefloch} B.,   {Maret} S.,  2006, \mn@doi [\aap] {10.1051/0004-6361:20054476}, \href {http://adsabs.harvard.edu/abs/2006A%26A...453..949P} {453, 949}

\bibitem[\protect\citeauthoryear{{Pazukhin}, {Zinchenko}, {Trofimova}, {Henkel}  \& {Semenov}}{{Pazukhin} et~al.}{2023}]{Pazukhin2023-Dhighhass}
{Pazukhin} A.~G.,  {Zinchenko} I.~I.,  {Trofimova} E.~A.,  {Henkel} C.,   {Semenov} D.~A.,  2023, \mn@doi [\mnras] {10.1093/mnras/stad2976}, \href {https://ui.adsabs.harvard.edu/abs/2023MNRAS.526.3673P} {526, 3673}

\bibitem[\protect\citeauthoryear{{Persson}, {J{\o}rgensen}, {van Dishoeck}  \& {Harsono}}{{Persson} et~al.}{2014}]{Persson2014}
{Persson} M.~V.,  {J{\o}rgensen} J.~K.,  {van Dishoeck} E.~F.,   {Harsono} D.,  2014, \mn@doi [\aap] {10.1051/0004-6361/201322845}, \href {https://ui.adsabs.harvard.edu/abs/2014A&A...563A..74P} {563, A74}

\bibitem[\protect\citeauthoryear{{Persson} et~al.,}{{Persson} et~al.}{2018}]{Persson2018}
{Persson} M.~V.,  et~al., 2018, \mn@doi [\aap] {10.1051/0004-6361/201731684}, \href {http://adsabs.harvard.edu/abs/2018A%26A...610A..54P} {610, A54}

\bibitem[\protect\citeauthoryear{{Podio} et~al.,}{{Podio} et~al.}{2021}]{Podio2021}
{Podio} L.,  et~al., 2021, \mn@doi [\aap] {10.1051/0004-6361/202038429}, \href {https://ui.adsabs.harvard.edu/abs/2021A&A...648A..45P} {648, A45}

\bibitem[\protect\citeauthoryear{{Podio} et~al.,}{{Podio} et~al.}{2024}]{Podio2024}
{Podio} L.,  et~al., 2024, \mn@doi [arXiv e-prints] {10.48550/arXiv.2407.04813}, \href {https://ui.adsabs.harvard.edu/abs/2024arXiv240704813P} {p. arXiv:2407.04813}

\bibitem[\protect\citeauthoryear{{Redaelli}, {Bizzocchi}, {Caselli}, {Sipil{\"a}}, {Lattanzi}, {Giuliano}  \& {Spezzano}}{{Redaelli} et~al.}{2019}]{Redaelli2019}
{Redaelli} E.,  {Bizzocchi} L.,  {Caselli} P.,  {Sipil{\"a}} O.,  {Lattanzi} V.,  {Giuliano} B.~M.,   {Spezzano} S.,  2019, \mn@doi [\aap] {10.1051/0004-6361/201935314}, \href {https://ui.adsabs.harvard.edu/abs/2019A&A...629A..15R} {629, A15}

\bibitem[\protect\citeauthoryear{{Riaz} \& {Thi}}{{Riaz} \& {Thi}}{2022}]{riaz2022-D2COinBDs}
{Riaz} B.,  {Thi} W.~F.,  2022, \mn@doi [\mnras] {10.1093/mnras/stac1573}, \href {https://ui.adsabs.harvard.edu/abs/2022MNRAS.514.3604R} {514, 3604}

\bibitem[\protect\citeauthoryear{{Riedel}, {Sipil{\"a}}, {Redaelli}, {Caselli}, {Vasyunin}, {Dulieu}  \& {Watanabe}}{{Riedel} et~al.}{2023}]{Riedel2023-Dmethanol}
{Riedel} W.,  {Sipil{\"a}} O.,  {Redaelli} E.,  {Caselli} P.,  {Vasyunin} A.~I.,  {Dulieu} F.,   {Watanabe} N.,  2023, \mn@doi [\aap] {10.1051/0004-6361/202245367}, \href {https://ui.adsabs.harvard.edu/abs/2023A&A...680A..87R} {680, A87}

\bibitem[\protect\citeauthoryear{{Rimola}, {Taquet}, {Ugliengo}, {Balucani}  \& {Ceccarelli}}{{Rimola} et~al.}{2014}]{Rimola2014}
{Rimola} A.,  {Taquet} V.,  {Ugliengo} P.,  {Balucani} N.,   {Ceccarelli} C.,  2014, \mn@doi [\aap] {10.1051/0004-6361/201424046}, \href {https://ui.adsabs.harvard.edu/abs/2014A%26A...572A..70R} {572, A70}

\bibitem[\protect\citeauthoryear{{Rivilla}, {Colzi}, {Fontani}, {Melosso}, {Caselli}, {Bizzocchi}, {Tamassia}  \& {Dore}}{{Rivilla} et~al.}{2020}]{Rivilla2020}
{Rivilla} V.~M.,  {Colzi} L.,  {Fontani} F.,  {Melosso} M.,  {Caselli} P.,  {Bizzocchi} L.,  {Tamassia} F.,   {Dore} L.,  2020, \mn@doi [\mnras] {10.1093/mnras/staa1616}, \href {https://ui.adsabs.harvard.edu/abs/2020MNRAS.496.1990R} {496, 1990}

\bibitem[\protect\citeauthoryear{{Roberts} \& {Millar}}{{Roberts} \& {Millar}}{2007}]{Roberts2007-D2COobs}
{Roberts} H.,  {Millar} T.~J.,  2007, \mn@doi [\aap] {10.1051/0004-6361:20066608}, \href {http://adsabs.harvard.edu/abs/2007A%26A...471..849R} {471, 849}

\bibitem[\protect\citeauthoryear{{Roberts}, {Herbst}  \& {Millar}}{{Roberts} et~al.}{2003}]{Roberts2003-multiplyD}
{Roberts} H.,  {Herbst} E.,   {Millar} T.~J.,  2003, \mn@doi [\apjl] {10.1086/376962}, \href {https://ui.adsabs.harvard.edu/abs/2003ApJ...591L..41R} {591, L41}

\bibitem[\protect\citeauthoryear{{Roueff}, {Tin{\'e}}, {Coudert}, {Pineau des For{\^e}ts}, {Falgarone}  \& {Gerin}}{{Roueff} et~al.}{2000}]{Roueff2000}
{Roueff} E.,  {Tin{\'e}} S.,  {Coudert} L.~H.,  {Pineau des For{\^e}ts} G.,  {Falgarone} E.,   {Gerin} M.,  2000, \aap, \href {https://ui.adsabs.harvard.edu/abs/2000A&A...354L..63R} {354, L63}

\bibitem[\protect\citeauthoryear{{Roueff}, {Parise}  \& {Herbst}}{{Roueff} et~al.}{2007}]{Roueff2007-warmDeuteration}
{Roueff} E.,  {Parise} B.,   {Herbst} E.,  2007, \mn@doi [\aap] {10.1051/0004-6361:20066531}, \href {https://ui.adsabs.harvard.edu/abs/2007A&A...464..245R} {464, 245}

\bibitem[\protect\citeauthoryear{{Sabatini} et~al.,}{{Sabatini} et~al.}{2020}]{Sabatini2020-H2D+}
{Sabatini} G.,  et~al., 2020, \mn@doi [\aap] {10.1051/0004-6361/202039010}, \href {https://ui.adsabs.harvard.edu/abs/2020A&A...644A..34S} {644, A34}

\bibitem[\protect\citeauthoryear{{Sabatini}, {Bovino}  \& {Redaelli}}{{Sabatini} et~al.}{2023}]{Sabatini2023-CR}
{Sabatini} G.,  {Bovino} S.,   {Redaelli} E.,  2023, \mn@doi [\apjl] {10.3847/2041-8213/acc940}, \href {https://ui.adsabs.harvard.edu/abs/2023ApJ...947L..18S} {947, L18}

\bibitem[\protect\citeauthoryear{{Santangelo} et~al.,}{{Santangelo} et~al.}{2015}]{Santangelo2015}
{Santangelo} G.,  et~al., 2015, \mn@doi [\aap] {10.1051/0004-6361/201526323}, \href {https://ui.adsabs.harvard.edu/abs/2015A&A...584A.126S} {584, A126}

\bibitem[\protect\citeauthoryear{{Song} \& {K{\"a}stner}}{{Song} \& {K{\"a}stner}}{2017}]{Song2017-H2CO&KIE}
{Song} L.,  {K{\"a}stner} J.,  2017, \mn@doi [\apj] {10.3847/1538-4357/aa943e}, \href {https://ui.adsabs.harvard.edu/abs/2017ApJ...850..118S} {850, 118}

\bibitem[\protect\citeauthoryear{{Su}, {Liu}, {Li}, {Lee}, {Hirano}, {Takakuwa}  \& {Hsieh}}{{Su} et~al.}{2019}]{Su2019}
{Su} Y.-N.,  {Liu} S.-Y.,  {Li} Z.-Y.,  {Lee} C.-F.,  {Hirano} N.,  {Takakuwa} S.,   {Hsieh} I.~T.,  2019, \mn@doi [\apj] {10.3847/1538-4357/ab4818}, \href {https://ui.adsabs.harvard.edu/abs/2019ApJ...885...98S} {885, 98}

\bibitem[\protect\citeauthoryear{{Taquet}, {Ceccarelli}  \& {Kahane}}{{Taquet} et~al.}{2012}]{Taquet2012-Deuteration}
{Taquet} V.,  {Ceccarelli} C.,   {Kahane} C.,  2012, \mn@doi [\apjl] {10.1088/2041-8205/748/1/L3}, \href {http://adsabs.harvard.edu/abs/2012ApJ...748L...3T} {748, L3}

\bibitem[\protect\citeauthoryear{{Taquet}, {L{\'o}pez-Sepulcre}, {Ceccarelli}, {Neri}, {Kahane}, {Coutens}  \& {Vastel}}{{Taquet} et~al.}{2013}]{Taquet2013-obs}
{Taquet} V.,  {L{\'o}pez-Sepulcre} A.,  {Ceccarelli} C.,  {Neri} R.,  {Kahane} C.,  {Coutens} A.,   {Vastel} C.,  2013, \mn@doi [\apjl] {10.1088/2041-8205/768/2/L29}, \href {https://ui.adsabs.harvard.edu/abs/2013ApJ...768L..29T} {768, L29}

\bibitem[\protect\citeauthoryear{{Taquet}, {L{\'o}pez-Sepulcre}, {Ceccarelli}, {Neri}, {Kahane}  \& {Charnley}}{{Taquet} et~al.}{2015}]{Taquet2015}
{Taquet} V.,  {L{\'o}pez-Sepulcre} A.,  {Ceccarelli} C.,  {Neri} R.,  {Kahane} C.,   {Charnley} S.~B.,  2015, \mn@doi [\apj] {10.1088/0004-637X/804/2/81}, \href {https://ui.adsabs.harvard.edu/abs/2015ApJ...804...81T} {804, 81}

\bibitem[\protect\citeauthoryear{{Taquet} et~al.,}{{Taquet} et~al.}{2019}]{Taquet2019}
{Taquet} V.,  et~al., 2019, \mn@doi [\aap] {10.1051/0004-6361/201936044}, \href {https://ui.adsabs.harvard.edu/abs/2019A&A...632A..19T} {632, A19}

\bibitem[\protect\citeauthoryear{{Taquet} et~al.,}{{Taquet} et~al.}{2020}]{Taquet2020}
{Taquet} V.,  et~al., 2020, \mn@doi [\aap] {10.1051/0004-6361/201937072}, \href {https://ui.adsabs.harvard.edu/abs/2020A&A...637A..63T} {637, A63}

\bibitem[\protect\citeauthoryear{{Tielens} \& {Hagen}}{{Tielens} \& {Hagen}}{1982}]{tielens1982}
{Tielens} A.~G.~G.~M.,  {Hagen} W.,  1982, \aap, \href {https://ui.adsabs.harvard.edu/abs/1982A&A...114..245T} {114, 245}

\bibitem[\protect\citeauthoryear{{Tsuge} \& {Watanabe}}{{Tsuge} \& {Watanabe}}{2023}]{TsugeWatanabe2023-RevPJAB}
{Tsuge} M.,  {Watanabe} N.,  2023, \mn@doi [Proceedings of the Japan Academy, Series B] {10.2183/pjab.99.008}, \href {https://ui.adsabs.harvard.edu/abs/2023PJAB...99..103T} {99, 103}

\bibitem[\protect\citeauthoryear{{Turner}}{{Turner}}{1990}]{Turner1990-D2CO}
{Turner} B.~E.,  1990, \mn@doi [\apjl] {10.1086/185840}, \href {https://ui.adsabs.harvard.edu/abs/1990ApJ...362L..29T} {362, L29}

\bibitem[\protect\citeauthoryear{{Vastel}, {Phillips}  \& {Yoshida}}{{Vastel} et~al.}{2004}]{Vastel2004-D2H+}
{Vastel} C.,  {Phillips} T.~G.,   {Yoshida} H.,  2004, \mn@doi [\apjl] {10.1086/421265}, \href {https://ui.adsabs.harvard.edu/abs/2004ApJ...606L.127V} {606, L127}

\bibitem[\protect\citeauthoryear{{Watanabe} \& {Kouchi}}{{Watanabe} \& {Kouchi}}{2002}]{Watanabe2002-methanol}
{Watanabe} N.,  {Kouchi} A.,  2002, \mn@doi [\apjl] {10.1086/341412}, \href {https://ui.adsabs.harvard.edu/abs/2002ApJ...571L.173W} {571, L173}

\bibitem[\protect\citeauthoryear{{Woon}}{{Woon}}{2002}]{Woon2002-COhydrogenation}
{Woon} D.~E.,  2002, \mn@doi [\apj] {10.1086/339279}, \href {https://ui.adsabs.harvard.edu/abs/2002ApJ...569..541W} {569, 541}

\bibitem[\protect\citeauthoryear{{Yamato}, {Furuya}, {Aikawa}, {Persson}, {Tobin}, {J{\o}rgensen}  \& {Kama}}{{Yamato} et~al.}{2022}]{Yamato2022-DammoniaIRAS4A}
{Yamato} Y.,  {Furuya} K.,  {Aikawa} Y.,  {Persson} M.~V.,  {Tobin} J.~J.,  {J{\o}rgensen} J.~K.,   {Kama} M.,  2022, \mn@doi [\apj] {10.3847/1538-4357/ac9ea5}, \href {https://ui.adsabs.harvard.edu/abs/2022ApJ...941...75Y} {941, 75}

\bibitem[\protect\citeauthoryear{{Yang} et~al.,}{{Yang} et~al.}{2021}]{Yang2021-peaches}
{Yang} Y.-L.,  et~al., 2021, \mn@doi [\apj] {10.3847/1538-4357/abdfd6}, \href {https://ui.adsabs.harvard.edu/abs/2021ApJ...910...20Y} {910, 20}

\bibitem[\protect\citeauthoryear{{Zahorecz}, {Jimenez-Serra}, {Testi}, {Immer}, {Fontani}, {Caselli}, {Wang}  \& {Toth}}{{Zahorecz} et~al.}{2017}]{Zahorecz2017}
{Zahorecz} S.,  {Jimenez-Serra} I.,  {Testi} L.,  {Immer} K.,  {Fontani} F.,  {Caselli} P.,  {Wang} K.,   {Toth} L.~V.,  2017, \mn@doi [\aap] {10.1051/0004-6361/201629792}, \href {https://ui.adsabs.harvard.edu/abs/2017A&A...602L...3Z} {602, L3}

\bibitem[\protect\citeauthoryear{{Zahorecz}, {Jimenez-Serra}, {Testi}, {Immer}, {Fontani}, {Caselli}, {Wang}  \& {Onishi}}{{Zahorecz} et~al.}{2021}]{Zahorecz2021-d2coHighMass}
{Zahorecz} S.,  {Jimenez-Serra} I.,  {Testi} L.,  {Immer} K.,  {Fontani} F.,  {Caselli} P.,  {Wang} K.,   {Onishi} T.,  2021, \mn@doi [\aap] {10.1051/0004-6361/201937012}, \href {https://ui.adsabs.harvard.edu/abs/2021A&A...653A..45Z} {653, A45}

\bibitem[\protect\citeauthoryear{{Zucker}, {Schlafly}, {Speagle}, {Green}, {Portillo}, {Finkbeiner}  \& {Goodman}}{{Zucker} et~al.}{2018}]{Zucker2018}
{Zucker} C.,  {Schlafly} E.~F.,  {Speagle} J.~S.,  {Green} G.~M.,  {Portillo} S.~K.~N.,  {Finkbeiner} D.~P.,   {Goodman} A.~A.,  2018, \mn@doi [\apj] {10.3847/1538-4357/aae97c}, \href {https://ui.adsabs.harvard.edu/abs/2018ApJ...869...83Z} {869, 83}

\bibitem[\protect\citeauthoryear{{van Gelder} et~al.,}{{van Gelder} et~al.}{2020}]{vanGelder2020}
{van Gelder} M.~L.,  et~al., 2020, \mn@doi [\aap] {10.1051/0004-6361/202037758}, \href {https://ui.adsabs.harvard.edu/abs/2020A&A...639A..87V} {639, A87}

\bibitem[\protect\citeauthoryear{{van der Tak}, {Schilke}, {M{\"u}ller}, {Lis}, {Phillips}, {Gerin}  \& {Roueff}}{{van der Tak} et~al.}{2002}]{vanderTak2002-ND3}
{van der Tak} F.~F.~S.,  {Schilke} P.,  {M{\"u}ller} H.~S.~P.,  {Lis} D.~C.,  {Phillips} T.~G.,  {Gerin} M.,   {Roueff} E.,  2002, \mn@doi [\aap] {10.1051/0004-6361:20020647}, \href {https://ui.adsabs.harvard.edu/abs/2002A&A...388L..53V} {388, L53}

\makeatother
\end{thebibliography}

\clearpage
\appendix

\section{Spectra}

\begin{figure*}
    \centering
    \includegraphics[width=0.98\textwidth]{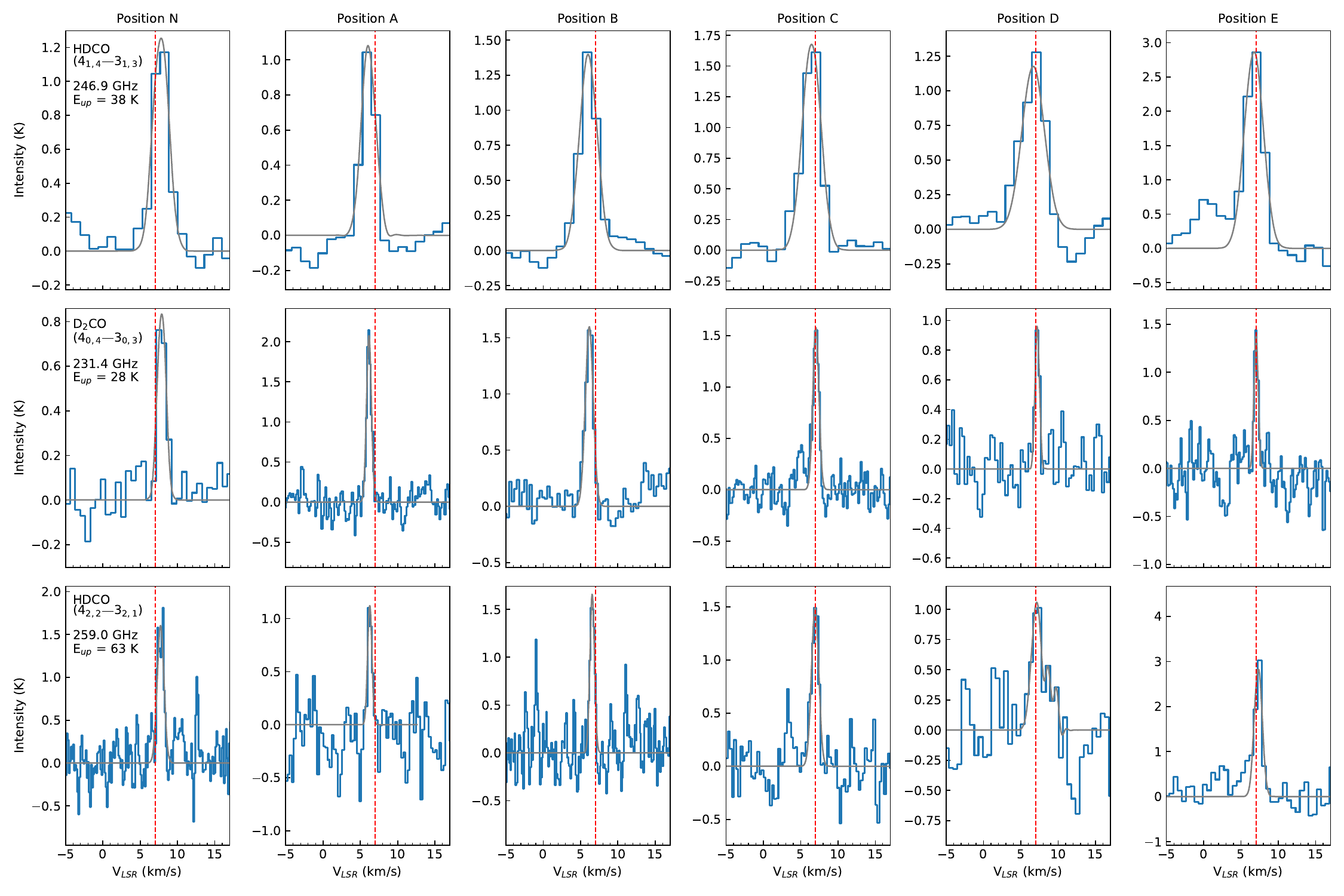} 

    \caption{Spectra of the HDCO(4$_{1,4}$--3$_{1,3}$), D$_{2}$CO(4$_{0,4}$--3$_{0,3}$) and HDCO(4$_{2,2}$--3$_{2,1}$) lines extracted at the positions N, A, B, C, D and E shown in Fig \ref{fig:hdco-d2co-ratio}. The positions are labelled on the top of the boxes in the first row. The line transition, frequency, and upper level energy are indicated on the upper-left of the panels in the first column. The red dashed line represents the systemic velocity, V$_{\rm{lsr}}$\,=\,7\,km\,s$^{-1}$. For better visualisation, the spectrum of D${2}$CO(4$_{0,4}$--3$_{0,3}$) at position B was smoothed once, and at position N was smoothed twice. Similarly, the HDCO(4$_{2,2}$--3$_{2,1}$) spectra at positions D and E were smoothed twice. Note that this smoothing was applied solely for this figure and was not used in the RD analysis. The Gaussian fits are shown in grey.}
    
   \label{fig:spectra}
\end{figure*}

In Fig. \ref{fig:spectra} we show the spectra of the HDCO(4$_{1,4}$--3$_{1,3}$), D$_{2}$CO(4$_{0,4}$--3$_{0,3}$) and HDCO(4$_{2,2}$--3$_{2,1}$) lines extracted at the positions shown in Fig. \ref{fig:hdco-d2co-ratio}. 

\section{LTE Rotational Diagram Analysis}
\label{appendix:lte}

\begin{figure*}
    \centering
    \includegraphics[width=\textwidth]{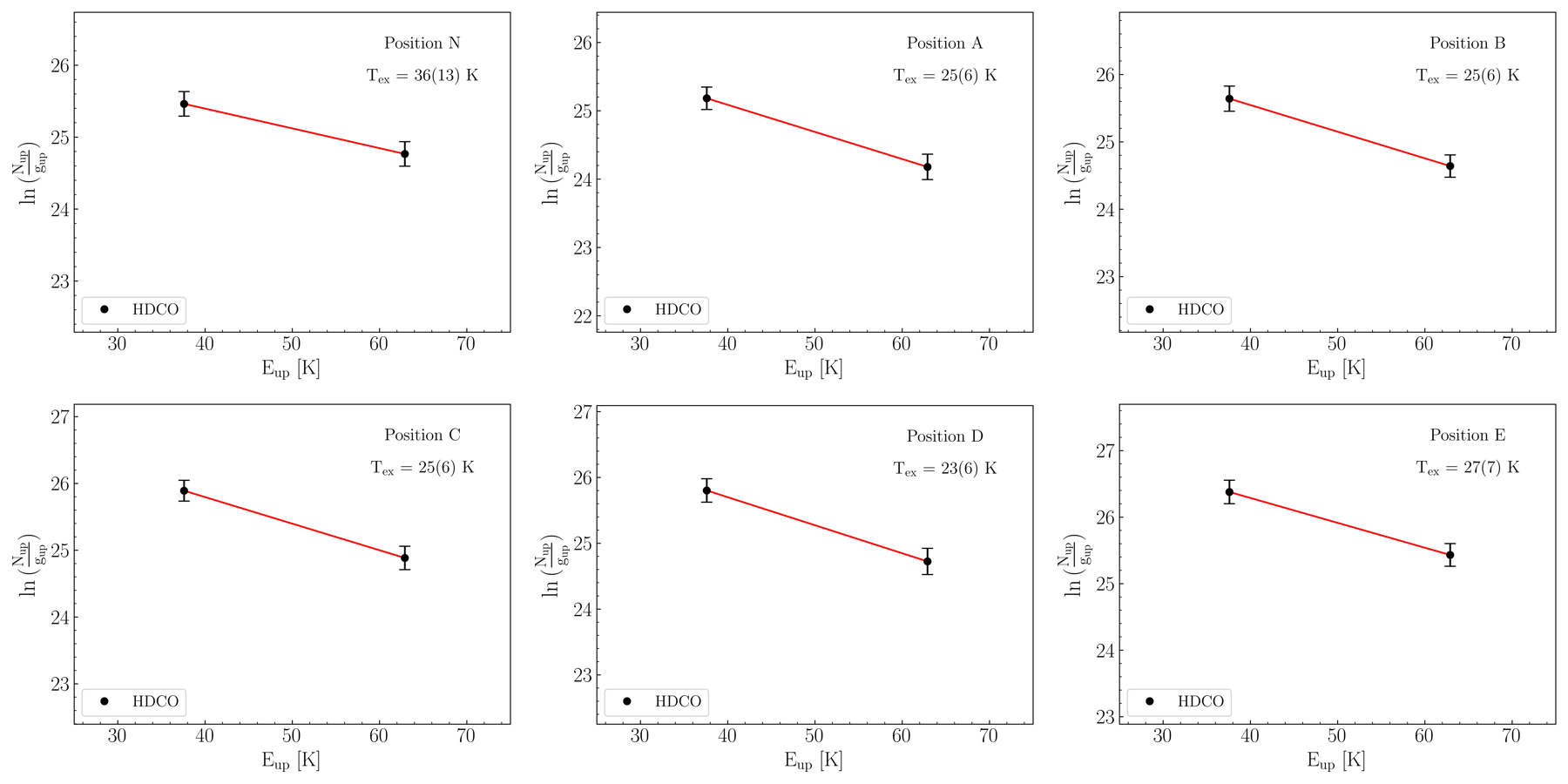} 
     
    \caption{Rotational diagrams for HDCO at the positions N, A, B, C, D and E shown in Fig \ref{fig:hdco-d2co-ratio}. The red lines correspond to the best fit to the data points.}
    
   \label{fig:rds}
\end{figure*}

To determine the temperature distribution along the IRAS\,4A1 cavity walls and the associated bow-shock structure, we performed a rotational diagram analysis of the HDCO lines at the specified positions (N, A, B, C, D, and E), as illustrated in Fig. \ref{fig:hdco-d2co-ratio}. Employing the local thermodynamic equilibrium (LTE) assumption and assuming optically thin lines, we fitted the data point using weighted linear regression with the Python function \textit{curve\_fit}, which accounts for measurement uncertainties (See Fig. \ref{fig:rds}). The line profiles were fitted with a Gaussian function with CLASS. Both the rms and calibration errors in the line intensities were considered in the analysis. In Fig. \ref{fig:rds} we show the Rotational Diagrams of HDCO at the same positions. In Table \ref{tab:line-int} we show the line fluxes and associated root mean square (rms) noise levels for the observed spectra at the same positions

\renewcommand{\arraystretch}{1.8}
\begin{table*}
    \centering
    \caption{Line fluxes and associated root mean square (rms) noise levels for the observed spectra at various positions along the southern cavity of IRAS\,4A1}
    \begin{tabular}{ccccccc}
    \hline
        \hline
            Position Name & $\mathrm{I}_{\mathrm{HDCO(4_{2,2}--3_{2,1})}}$  & $\mathrm{rms\,}_{\mathrm{HDCO(4_{2,2}--3_{2,1})}}$ & $\mathrm{I}_{\mathrm{HDCO(4_{1,4}--3_{1,3})}}$ & $\mathrm{rms\,}_{\mathrm{HDCO(4_{1,4}--3_{1,3})}}$  & $\mathrm{I}_{\mathrm{D_2CO(4_{0,4}--3_{0,3})}}$ & $\mathrm{rms\,}_{\mathrm{D_2CO(4_{0,4}--3_{0,3})}}$ \\

         & (K km s$^{-1}$)& (K) & (K km s$^{-1}$) & (K) & (K km s$^{-1}$) & (K) \\
         \hline  

        N & $1.4 \pm 0.2$  & 0.23 & $3.4 \pm 0.3 $ & 0.05 & $1.5 \pm 0.2$ & 0.10 \\
        A & $0.8 \pm 0.2$  & 0.19 &  $2.6 \pm 0.2$ & 0.06 & $1.7 \pm 0.1$ & 0.14 \\
        B & $1.3 \pm 0.1$  & 0.27 &  $4.1 \pm 0.4$ & 0.07 & $2.1 \pm 0.2$ & 0.12 \\
        C & $1.6 \pm 0.2$  & 0.16 &  $5.3 \pm 0.3$ & 0.06 &$1.6 \pm 0.1$ & 0.12 \\
        D & $1.4 \pm 0.3$  & 0.31 &  $4.8 \pm 0.4$ &0.12 & $0.7 \pm 0.1$ & 0.20 \\
        E & $2.8 \pm 0.2$  & 0.30 & $8.6 \pm 0.6$ & 0.06 & $1.0 \pm 0.2$ & 0.21 \\
        
         \hline  
    \end{tabular}
    \label{tab:line-int}
\end{table*}
 
\bsp	
\label{lastpage}
\end{document}